# Development of a VR tool to study pedestrian route and exit choice behaviour in a multi-story building

YAN FENG[1*], DORINE DUIVES[1], SERGE HOOGENDOORN[1]

[1]Department of Transport & Planning, Delft University of Technology, 2628 CN, Delft, The Netherlands

**Abstract:** Although route and exit choice in complex buildings are important aspects of pedestrian behaviour, studies predominantly investigated pedestrian movement in a single level. This paper presents an innovative VR tool that was designed to investigate pedestrian route and exit choice in a multi-story building. This tool supports free navigation and collects pedestrian walking trajectories, head movements and gaze points automatically. An experiment was conducted to evaluate the VR tool from objective standpoints (i.e., pedestrian behaviour) and subjective standpoints (i.e., the feeling of presence, system usability, simulation sickness). The results show that the VR tool allows for accurate collection of pedestrian behavioural data in the complex building. Moreover, the results of the questionnaire report high realism of the virtual environment, high immersive feeling, high usability, and low simulator sickness. This paper contributes by showcasing an innovative approach of applying VR technologies to study pedestrian behaviour in complex and realistic environments.

## 1 INTRODUCTION

While walking in a building, pedestrians constantly choose between a number of routes to reach their destination, which is referred to as route choice behaviour [1]. On their way out, pedestrians are furthermore required to choose an exit. This exit choice behaviour features the choice of one exit within a set of alternative exits to leave certain places [2]. Many disciplines, such as architecture, fire safety engineering, and civil engineering, require a thorough understanding of pedestrian route and exit choice in buildings in order to ensure pedestrian safety and design comfortable buildings [3].

Traditionally, field experiments have been widely applied to investigate pedestrian route and exit choice behaviour. Field experiments collect pedestrian route and exit behaviour data in real-life conditions under uncontrolled (e.g., [4–7]) or controlled conditions (e.g., [8–11]). Digital equipment (e.g., camera) is usually used to record pedestrian behaviour in specific situations or particular locations. However, it is difficult to control external factors in field experiments under uncontrolled conditions [12]. Moreover, the raw data captured during a field experiment cannot be analysed directly, often the data still need to be extracted from a video recording afterwards. Consequently, it requires large labour and monetary investments to perform field experiments, while the data is often not accurate and reliable enough to perform intricate data analysis. Furthermore, controlled experiments to study pedestrian behaviour in risky situations are often restricted by ethical considerations featuring the mental and physical health of participants [13].

In order to overcome these limitations, researchers have attempted to use Virtual Reality (VR) technologies to study pedestrian route and exit choice, especially during evacuations (e.g., [14–18]). Compared to field experiments, VR provides possibilities to obtain complete experimental control and collect accurate behavioural data automatically [12]. Moreover, VR allows participants to be immersed in dangerous environments without the risk of facing real danger.

[*] Corresponding author: Yan Feng, Email: y.feng@tudelft.nl

Existing studies have predominantly investigated pedestrian route and exit behaviour in simplified virtual environments, mostly a single level of a building and in particular pedestrian movements in the horizontal level have been studied (e.g.,[14,15,19]). Moreover, some studies with VR simulators (e.g., desktop-based VR) recorded issues such as lack of natural movements, missing details of real-life situations, which might lead to the unrealistic perception of the virtual environment. Consequently, studies that use VR to collect comprehensive pedestrian route and exit choice behaviour in immersive, realistic and complex buildings are very rare.

This study aims to develop a VR research tool that addresses these gaps and unlocks the potential of VR technologies for the study of pedestrian route and exit choice behaviour in complex multi-story buildings. Maya and Unreal Engine 4 have been used to develop this VR tool. This paper details the development process of the VR tool and describes a preliminary experiment using this VR tool. The experiment was conducted to demonstrate the ability of this VR tool, and evaluate the usability and realism of the developed virtual environment.

The rest of the paper is organised as follows. Section 2 summarises different methods to collect pedestrian route and exit choice data in buildings. Section 3 presents the functional requirements of the VR research tool. Accordingly, section 4 details the developing process of the VR research tool. Section 5 details the experiment method applying this VR tool. The first results of this experiment are discussed in section 6. The paper ends with conclusions and future research.

## 2 RELATED WORK

People need to find their way through buildings while moving from one location to another. This behavioural process may be as easy as moving from one room to another or as difficult as trying to escape a building that is on fire [20]. Up to this moment, predominantly two experimental methods have been used to study pedestrian route and exit choice behaviour in buildings, namely field experiments and VR experiments. This section provides a brief overview of the work featuring these experimental methods related to pedestrian route and exit choice behaviour.

### 2.1 Field experiments

Field experiments have been wildly applied to study pedestrian route and exit behaviour, both in normal and emergency conditions. The major advantage of field experiments is that pedestrians walk in a real-life environment and are most likely to behave naturally. Often, cameras are used to record the movement and choice behaviour of pedestrians. Pedestrian route and exit choice behaviour have been investigated by means of field experiments in schools, universities, theatres, hospitals, tunnels [4–7,21–25]. These studies have illustrated that field experiment is a valuable experimental method to study pedestrian movement and choice behaviour.

However, in these field experiments, the experimental scenarios are generally difficult to control due to the complexity of most pedestrian infrastructures and natural variation of human behaviour in such environments [26]. Consequently, it has proven difficult to capture detailed data to characterise pedestrian behaviour [27] and challenging to isolate the effect of one explicit variable on pedestrian behaviour within a complex context. In the examples where an evacuation was studied (e.g., Fridolf et al., 2013; Heliövaara et al., 2012), the scenarios were not completely realistic due to ethical and financial constraints. Meanwhile, almost all studies into pedestrian route and exit behaviour have limited themselves to investigate pedestrian movement in the horizontal levels [9], most likely to curb the complexity of the experimental setup. Consequently, literature applying field experiments does not capture the complexity and difficulty of pedestrian movements in a multi-story building with both horizontal and vertical movements in a long distance.

### 2.2 VR laboratory experiments

Due to the above-mentioned limitations of field experiments, researchers have explored VR as an innovative experimental approach to study pedestrian behaviour. Using VR technologies, it is possible to automatically collect detailed behavioural data in immersive environments and analyse precisely how specific controlled factors influence pedestrian behaviour. Another benefit of VR is that it can be used to create environments that there are either not likely to encounter in real-life or scenarios which are too dangerous to expose a participant due to the health risks, for example, fire, smoke and terrorist attacks.

Existing research has investigated pedestrian route and exit behaviour in normal conditions or evacuations. This line of research is of two types, namely pedestrian's choice of route and exit (e.g.,[8,16,28–30]) and the impact of external factors on route and exit choice (e.g., [17,31–36]).

These studies illustrated that VR is a safe, engaging and appealing approach to study pedestrian route and exit choice behaviour. Moreover, these studies also provided some valuable insights regarding the optimal development and usage of VR technologies for pedestrian research. Firstly, the realism level of the virtual environment can impact on the accuracy of the behavioural data [37]. Existing studies have predominantly investigated simplified environments, studies featuring pedestrian route and exit choice behaviour in complex buildings are still rare. In order to collect more accurate pedestrian behaviour data, the developed virtual environments should represent realistic and complex real-life scenarios. Moreover, it is important to design realistic soundscapes to envelop the user in the ongoing situation, especially during emergencies [38]. Secondly, to establish the validity of the VR system, it is important to test whether the results from VR experiments aligns with the actual behaviours of pedestrian in the real world. Only a few studies have attempted to validate their results via comparing pedestrian route and exit choice behaviour in VR and real-life scenarios (e.g., [16,30,39,40]). Thirdly, the literature suggests that more immersive virtual environments help participants behave closely to their behaviour in reality and consequently promise improved validity [41,42]. Compare to desktop VR, highly immersive VR systems, such as HMD and CAVE systems can provide more or full immersion for participants with more realistic feelings (e.g., [17,33,36,43–47]). Moreover, VR systems equipped with motion tracking devices (e.g., eye, head tracking devices) can more precisely measure visual attention and help researchers to gain a deeper understanding of how pedestrian interact with the environment. Lastly, the VR system should be easy to understand, use and interact so that it reduces the possibilities for participants experiencing simulation sickness [48,49]

To summarise, although pedestrian route and exit choice behaviour have been widely studied in field experiments or VR experiments, few studies have attempted to collect pedestrian behavioural data with both horizontal and vertical movements in realistic and complex environments. Pedestrian route and exit choice is affected by the layout of the architectural setting and the quality of the environmental information [20]. Since the complexity and difficulty of pedestrian movements in complex environments are very different [50], findings pertaining to simplified environments cannot be directly generalised to complex buildings. As a result, given the opportunities provided by VR, there is a strong need for VR tools that can create realistic and complex scenarios to study pedestrian route and exit choice in multi-story buildings, which includes both vertical and horizontal movements.

## 3 FUNCTIONAL REQUIREMENTS OF THE VR RESEARCH TOOL

The aim of the study is to develop a VR tool to study pedestrian route and exit choice behaviour in a multi-story building. Based on the aim of the study and review of previous studies pertaining to experimental designs to study pedestrian behaviour, we have identified five main requirements for the development of the new VR research tool.

Firstly, in order to study pedestrian route choices and exit choices across horizontal and vertical levels, the VR tool needs to allow users to perform wayfinding in multi-story buildings. Thus, the virtual environment is required to represent a complex building including multiple floors that are connected by means of staircases. Meanwhile, minimum of two set of route choices on both horizontal and vertical level is required.

Secondly, in order to allow for the validation of the VR tool, the virtual environment should feature a scenario that can be reproduced in reality, including all its intricacies. That is, the visualisation of the geometry, colour and texture in the virtual environment should be realistic to represent the real-world experience. Moreover, the general feeling of the environment should be similar as well. Thus, the details of the environment (e.g., signage, soundscapes) should be similar to a real-world experience.

Thirdly, in order to ensure the validity of the VR tool, the interaction between users and the virtual environment should be natural so that the participant can behave and react to events (e.g., evacuation) similarly to their real-life behaviour. To achieve the most natural response possible, the virtual environment needs to be immersive and interactive. To achieve full immersion, the VR tool should integrate natural navigation and realistic soundscapes.

Fourthly, the VR research tool is particularly designed to perform experiments. Thus, a major requirement of the VR tool is its ability to collect pedestrian behaviour data. In particular, the VR tool should be able to track participant's movements, choice and gazing behaviour (e.g., walking trajectory, timestamp, head rotation, gaze point). The VR tool should be able to repeatedly perform (almost) identical experiments with varying participants. Therefore, it should support slightly alter of the experimental setup per participant, while ensuring an as similar as possible experience. For instance, the viewpoint of participants should be able to be adjusted according to their height.

Lastly, the VR tool should be easy and comfortable to use for the participants and the researcher. This requirement does not only relate to the participants' ability to quickly learn how to use and interact with the VR tool but also the participants' mental and physical load of using the VR tool should not cause simulation sickness. Moreover, also the interface between the researcher and the VR environment should be relatively well-balanced in order to ease the operation of VR experiments.

The following sections address how we achieve the above-mentioned requirements and develop the VR tool.

## 4 DEVELOPMENT OF THE VR RESEARCH TOOL

To provide new opportunities for studying pedestrian route and exit choice in complex buildings, a new VR research tool has been developed. The development process of this VR tool considers three steps, namely 1) choice of the virtual environment, 2) construction of the virtual environment and 3) implementation of the interactive elements in the virtual environment. This section details the steps one by one.

### 4.1 Virtual environment layout

The aim is to study pedestrians' horizontal and vertical movement behaviour in complex scenarios which better reflect the actual situations people experience. Thus, the experimental environment should ideally be a building with multiple floors that enable pedestrians to choose between multiple routes and exit choices. Moreover, in the later stage of this research project, the authors aim to validate the results generated from the VR tool. Thus it should be possible to recreate the VR scenario in a real-life setting. Consequently, the choice has been made to recreate an existing real-life multi-story building in VR at a very high level of detail.

In this case, the building of Civil Engineering and Geoscience Faculty of Delft University of Technology has been chosen as the real-world benchmark of the virtual environment. This faculty building consists of seven floors; most of which feature two parallel running hallways, elevators and staircases that run through all levels of the building. Students mainly occupy the lower two floors and the top floor of the faculty building, while the faculty staff have their offices on the second to fifth floors.

To limit the difficulty of assignment performance in the virtual environment, the three intermediate floors of the building (the second, third and fourth floor) are chosen as the experimental area (Figure 1). This is the smallest number of floors required to test pedestrian route and exit behaviour featuring both horizontal and vertical levels. The layout of all three floors is quite similar, except for one additional space in between the two corridors on the second floor. That is, each floor has eight small corridors connecting the two main corridors. Besides that, each floor has five staircases (green) and five elevators (blue).

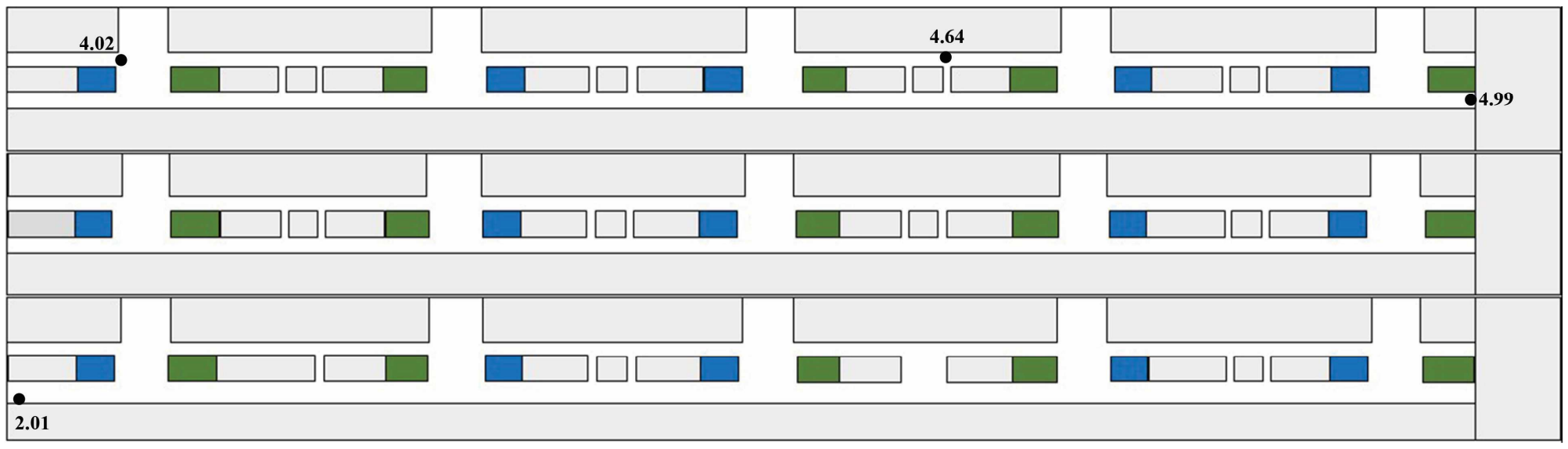

Figure 1: Floorplan of the Faculty of Civil Engineering and Geosciences

### 4.2 Construction of the virtual environment

The construction of the virtual environment featured two steps, namely the development of a 3D model of the building and the creation of the virtual environment. Firstly, the 3D model of the building was developed. Secondly, the virtual environment was developed based on the 3D model.

The first step was logging the details of the existing building by means of a pre-existing outdated 3D model of the building, site visits and photographs taken at the building by the researchers. Afterwards, the building was modelled in 3D using the combined information from different sources featuring the main characteristics of the building. The overall geometry for the 3D model was created using Maya. Here, three floors were created separately. The fourth floor was first built, and the second and third floors were built using the fourth floor as a base model because the main geometry of each floor is quite similar. Lastly, an exit floor was developed which connects to the second floor of the building. There were five exits located on the exit floor, namely Exit A, B, C, D and E. The main entrance of the building is Exit C. Figure 2 shows an overview of the comprehensive virtual building.

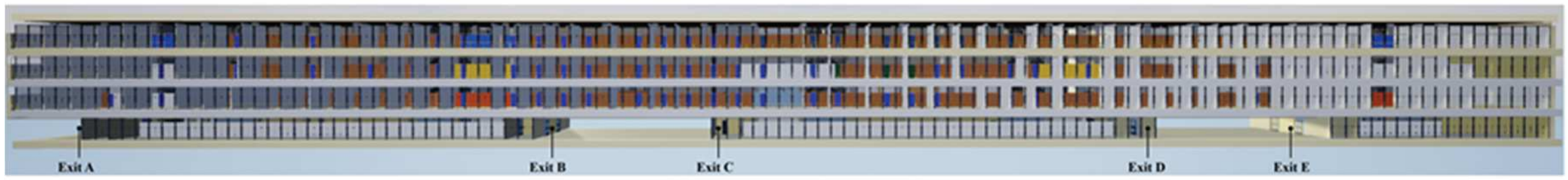

Figure 2: The overview of the virtual building

Once the overarching geometry (i.e., the internal layout of the building, walls, escalators, staircases) was finished, additional environmental elements were added to the 3D model to improve the accuracy of the building's representation and increase its realism. Four types of features were identified by Weisman (1981) as four classes of environmental variables that influence pedestrian route and exit choice behaviour within built environments, namely (a) visual access which provides views that one can see other parts of the building from a given location (e.g., glass windows), (b) architectural differentiation, which is the difference of objects in the building with respect to size, colour, location, etc. (e.g., chairs, cabinets, tables), (c) signs to provide identification or directional information (e.g., evacuation signs, exit signs, room numbers), and (d) plan configuration of the building (e.g., floor plan) [9,52]. These types of features were modelled in the virtual building in a way that they, as much as possible, resemble the current details in the building and placed in their original position. Figure 3 shows four examples of above-mentioned features that were added to the virtual environment.

The second step was creating the virtual environment. Using the 3D model of the building, the virtual environment was created in a game development engine, being Unreal  Engine 4 (UE4). UE4 is an open and widely used game engine developed by Epic Games (Epic Games, 2019). The UE4 was chosen for developing the complex virtual environment because it provides all the tools required to produce a high-quality virtual environment, and its built-in support for VR development makes it easy to work with VR hardware (e.g., HTC Vive and Oculus Rift). Furthermore, UE4 builds game levels which are texture-baked, compiled binaries that the game engine can efficiently process when running the game [53].

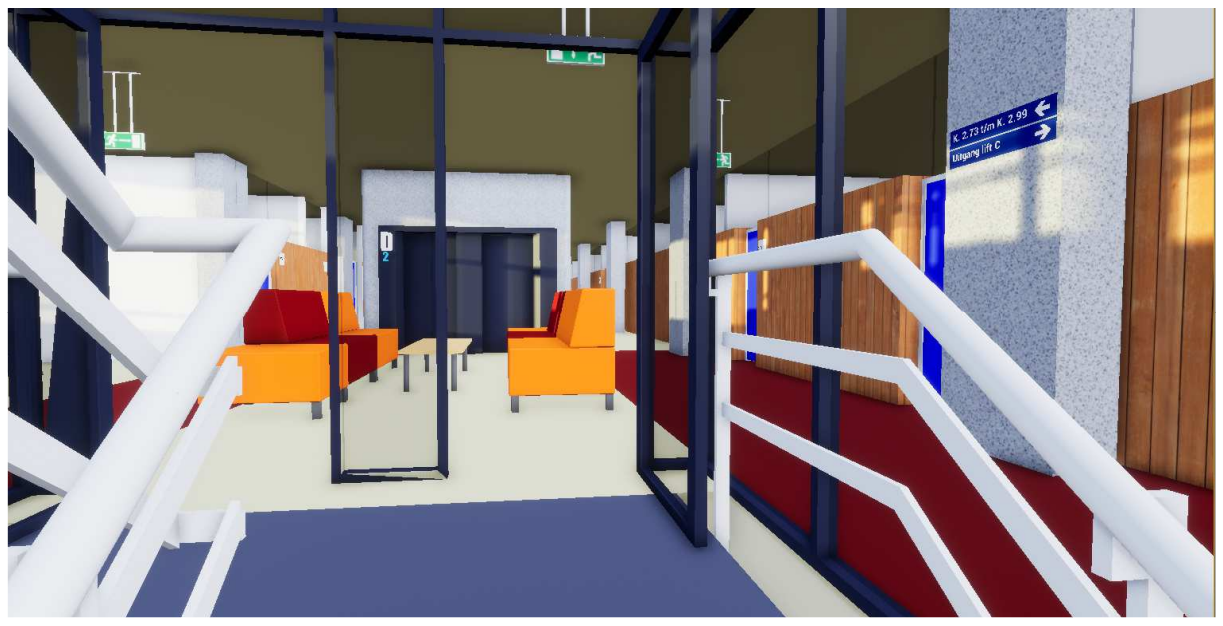

a. glasses window

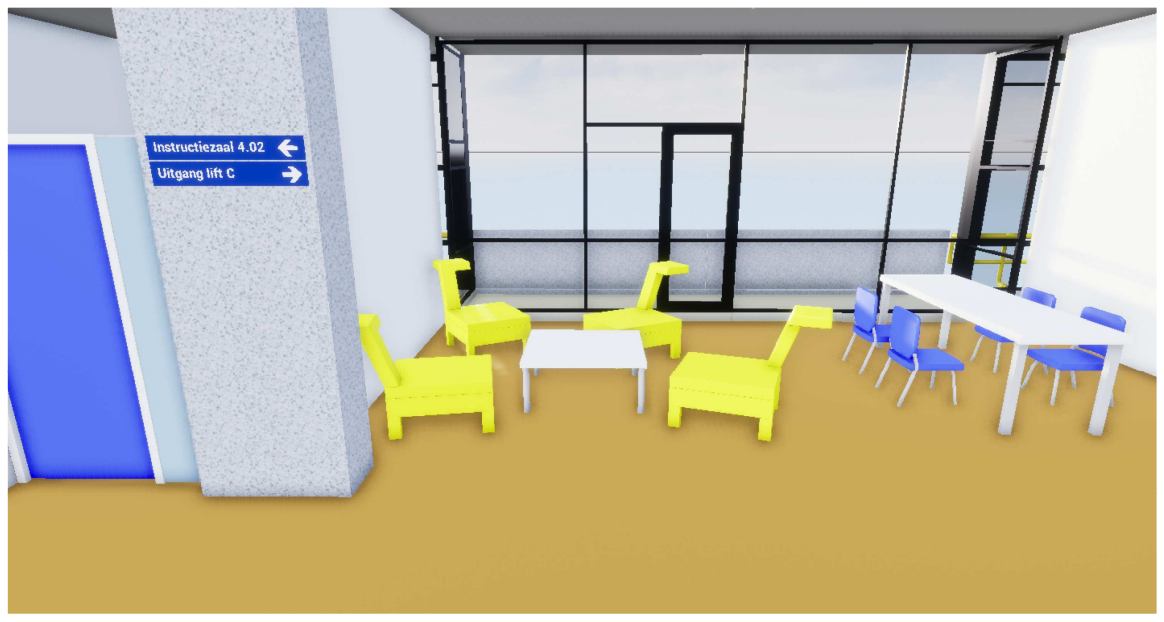

b. different chairs and tables

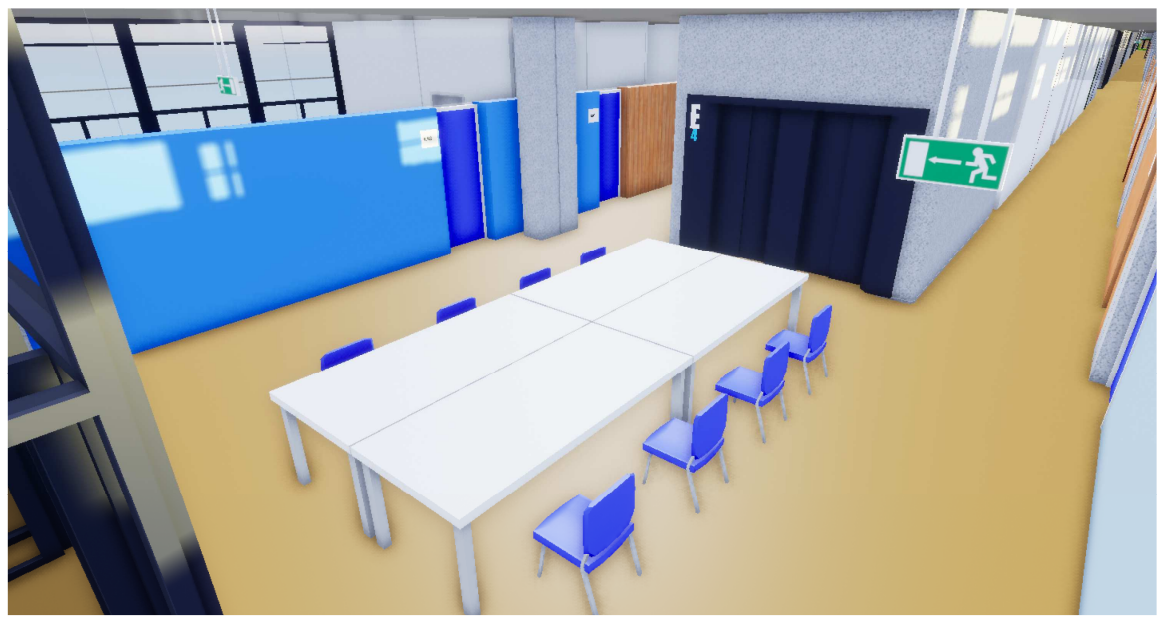

c. evacuation sign

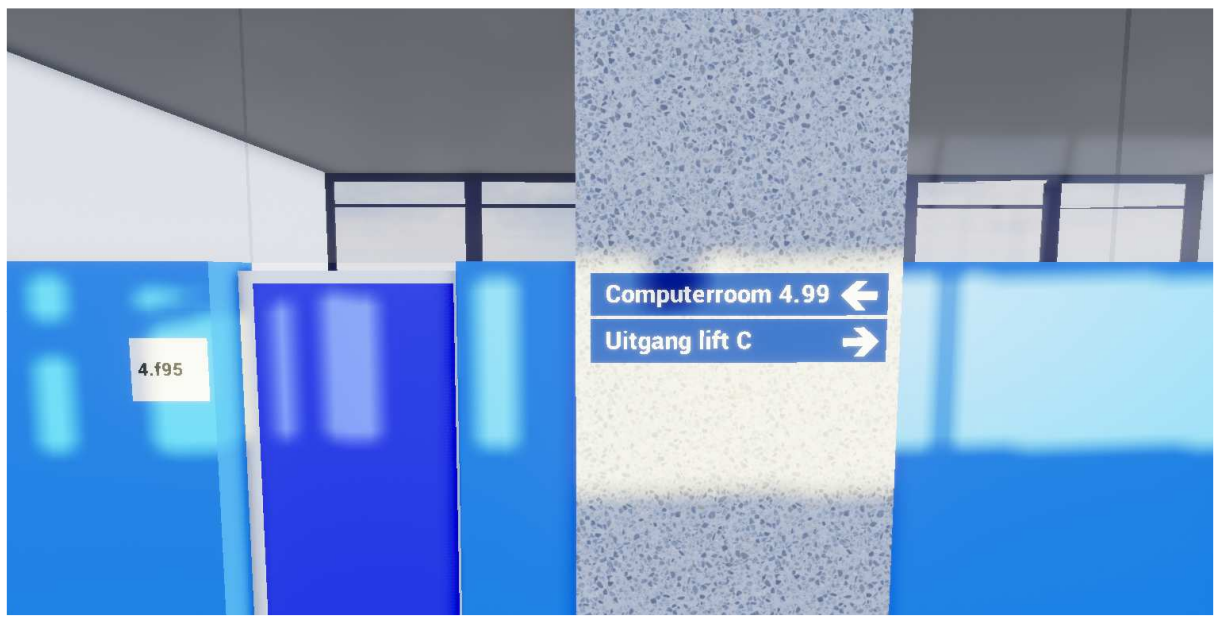

d. floor plan

Figure 3: Samples of four types of features added to the virtual environment

The 3D model was imported from Maya to UE4 using the FBX file format, which is directly readable by UE4. This static model in UE4, was accordingly used to render the virtual environment. Rendering effects include, for instance, textures, shadow, lighting, reflection, transparency. Deferred Renderer was selected as the rendering solution for the virtual environment, which is the default setting of UE4. Compared to forward rendering, which lighting has to be calculated for each vertex or pixel, deferred renderer is able to only run a single fragment shader for each render targets, which optimises complex scenes with a number of lights.

The colours and textures of objects in the virtual environment resemble those of the objects in the current faculty building as much as possible. In the virtual building, the corridors feature a mixture of yellow linoleum, coloured plaster walls (e.g., yellow, blue, orange), wooden panelling, rough concrete pillars and walls, and glass walls (Figure 4). Special attention was paid to ensure the correct representation of these four materials, given that they severely influence pedestrian's experience in the corridors and visibility of the stairs. Figure 5 shows one example of the final rendering of the virtual environment and the real-world view.

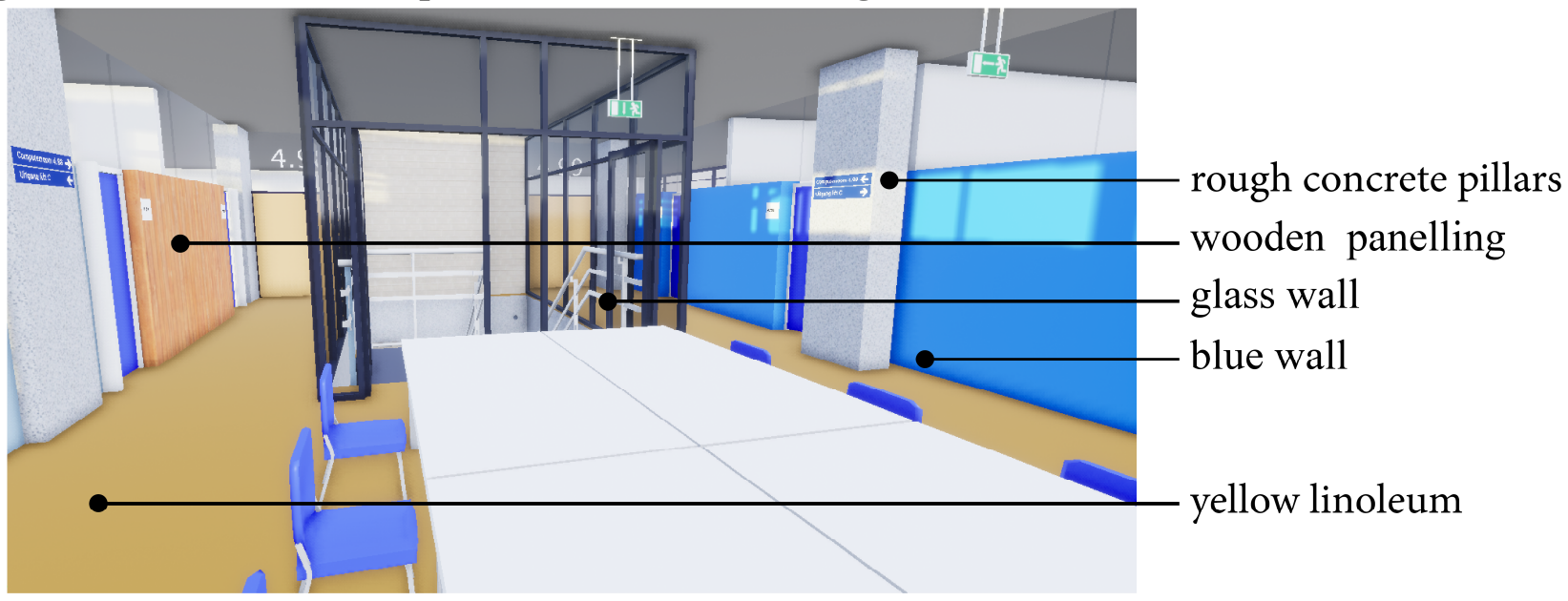

Figure 4: Illustration of one corridor in the virtual building

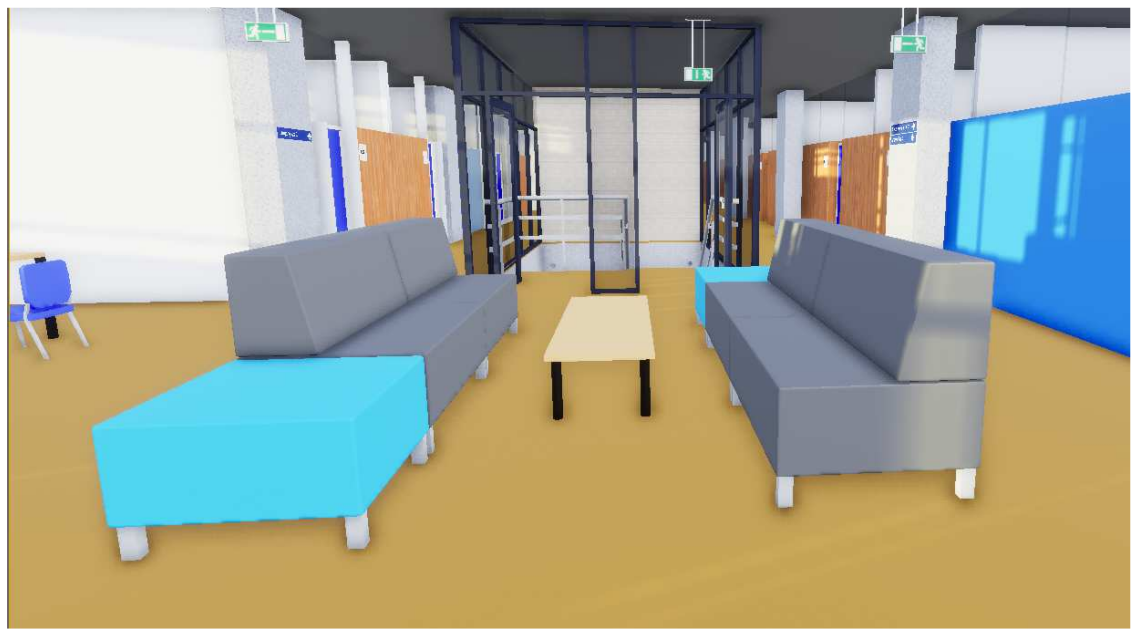

a. virtual building

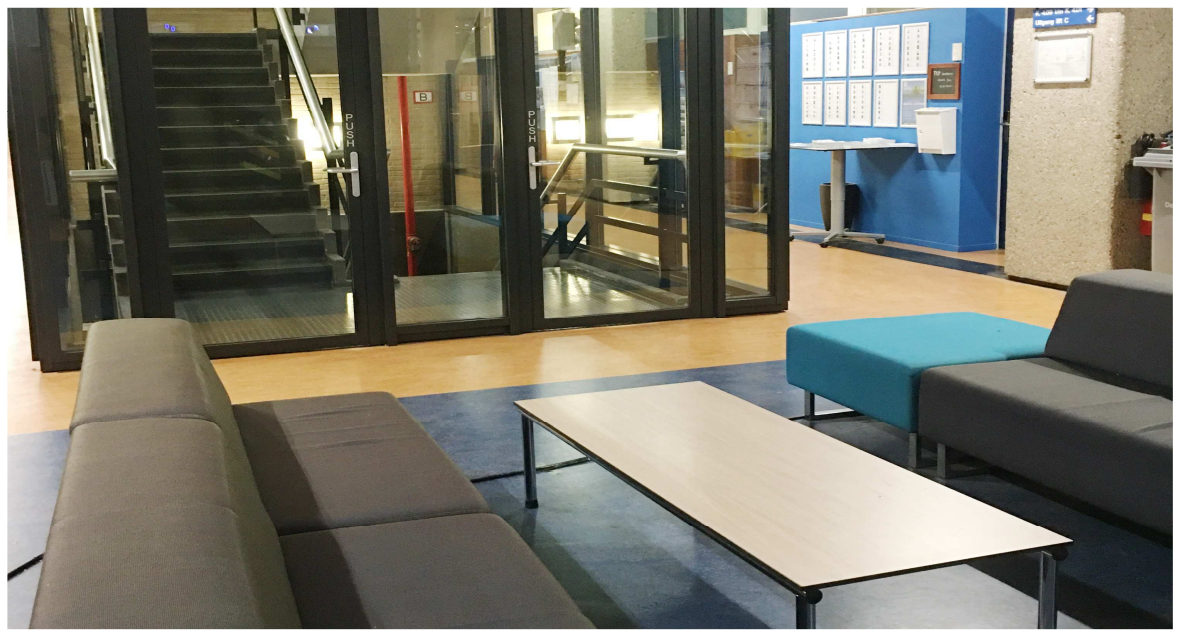

b. real world view

Figure 5: Screenshots of (a) the virtual building and (b) the real world view

### 4.3 Implementation of interaction elements

In addition to constructing a realistic virtual environment, the VR tool should support user interaction and provide an immersive environment to perform experiments. Thus it is necessary to integrate navigation, viewpoint, trigger, soundscape, and data recording. This section details the integration of these elements to the VR research tool using UE4.

1. *Navigation and locomotion*

In order to enable free navigation in the virtual building, similar to how pedestrians move freely in a real-life building, a combination of open world navigation solution and steering locomotion was implemented. This combination of both solutions reduces the change that users would experience motion sickness.

The open-world solution was achieved via the implementation of Navigation Mesh (NavMesh) in UE4, which defines the area users are able to walk in the building in order to explore the virtual environment (Figure 6). The NavMesh was only built within the walkable space in corridors, while the spaces of offices, elevators or obstacles (e.g., walls, furniture, objects) were not included. This mesh was adopted because of two reasons. First, it protects users from running into walls or other obstacles in the virtual building to initiate unrealistic experience. Second, it is on the authors' assumption that when people are required to evacuate from the building, office's doors and elevators would be inaccessible and unreachable.

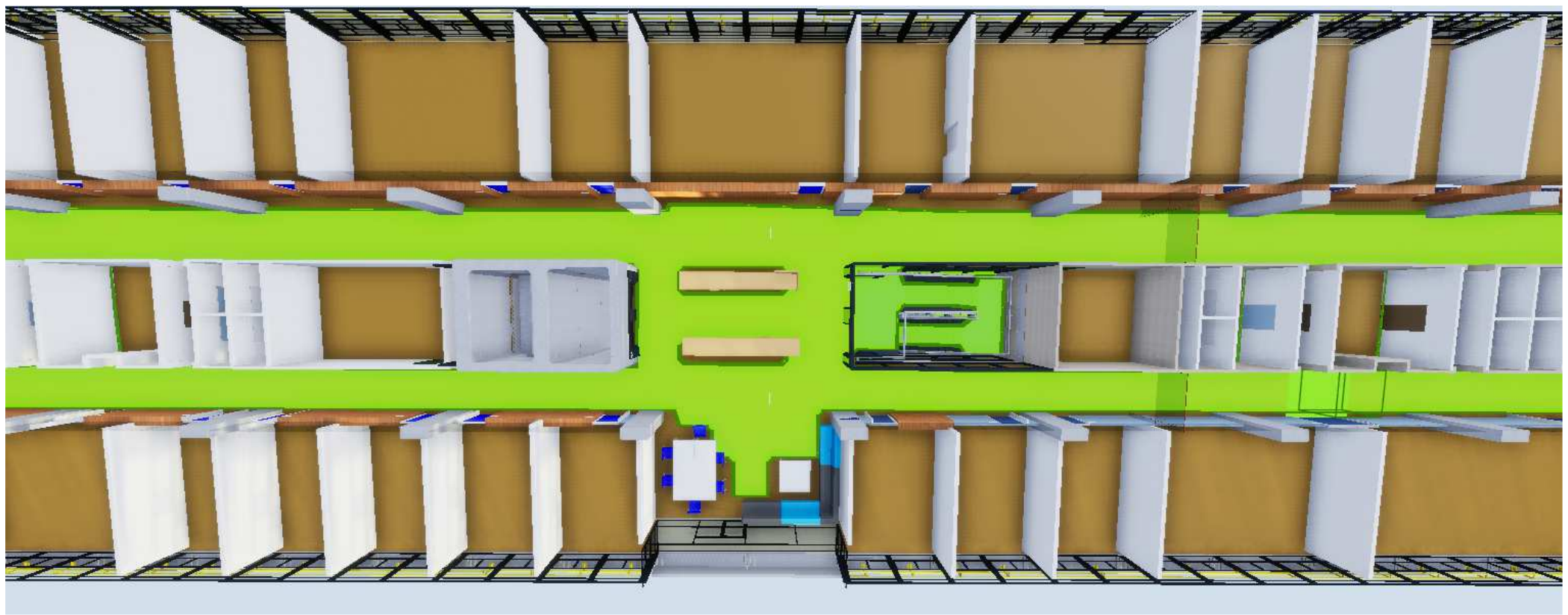

Figure 6: One example of implemented Navigation Mesh, indicated by green colour

Steering locomotion provides continuous movement and rotation in virtual space using a hand controller. This particular locomotion method allows for effective exploration and interaction with the virtual environment. In the prototype tests, we also found the steering locomotion generate less motion sickness compared to teleportation method. Meanwhile, the lack of continuous motion during teleportation might weaken presence and alert users that they are in a virtual environment.

Through our test, the maximum movement speed in the virtual environment has been limited to 140 cm/s to ensure that participants in the virtual building have, as much as possible, the same walking pace of the spatial layout as pedestrians have in real life (e.g., [54]). Meanwhile, our test showed that the speed limit minimises the motion sickness while participants moving in the virtual environment.

The direction of participant's movement in the virtual environment is controlled by their head rotations towards the direction they want to walk. This solution reduces the sickness as the rotations in the virtual and physical environments are the same.

*2. Viewpoint and avatar*

In UE4, participants' viewpoints are represented by a camera. Participants view the environment in the first-person perspective. Upon starting the simulation, the camera is located at a pre-defined start point. Once tracking is established and the user is on the starting position, the viewpoint is automatically calibrated to the actual height of the participant. As such, the user's vantage point in the virtual environment is similar to the user's 'normal' vantage point in real life.

Literature suggests pedestrian route and exit choice behaviour is affected by two major physical factors: the layout of the setting and the quality of the environmental information [9]. When evaluating the developed virtual environment, we were primarily interested in how pedestrians interact with the environment. Thus no other avatars were added in the environment at this stage.

*3. Trigger*

The virtual environment is designed in a way that participants can perform wayfinding assignments through the building. Thus, at various specific locations in the building, triggers were placed in order to present information messages to participants. When participants enter theses specific locations in the building, information messages would be triggered. These messages appear on the VR glasses screen and present a new (wayfinding) assignment to the participant. The virtual environment contains a sequence of different triggers. In case if participants enter one of the triggers' location without finishing the last assignment, the next trigger would not be activated.

*4. Soundscape*

In order to investigate pedestrian route and exit choice behaviour during evacuation, a scenario of evacuation drill was also stimulated by means of the VR tool. Thus, a 3D soundscape with realistic alarm sounds was incorporated that is also used during official evacuations at the faculty building of Civil Engineering and Geosciences. The alarm sound consists of a female voice that repeated the following statement: "Attention, please leave the building using the emergency exits as indicated. Do not use the elevators.".

*5. Data recording*

In order to function as a research tool, the VR tool needs to be able to record specific data points for later analysis. The position of the participant inside the virtual environment is obtained via the tracking system. All the parameters related to viewpoint's locations, such as positional data (x, y, z), head rotations (yaw, roll, pitch), gaze points, timestamps are recorded in milliseconds. All information is saved in separate CSV files per participant, which can be easily interpreted using data analytic toolboxes such as Python, R and Matlab. It can also be visualised in the virtual building using the built-in playback system to review what happened at a specific location or timestamp. For instance, Figure 7 shows the distribution of one user's walking trajectories (lines) and gaze points (dots) in the virtual building.

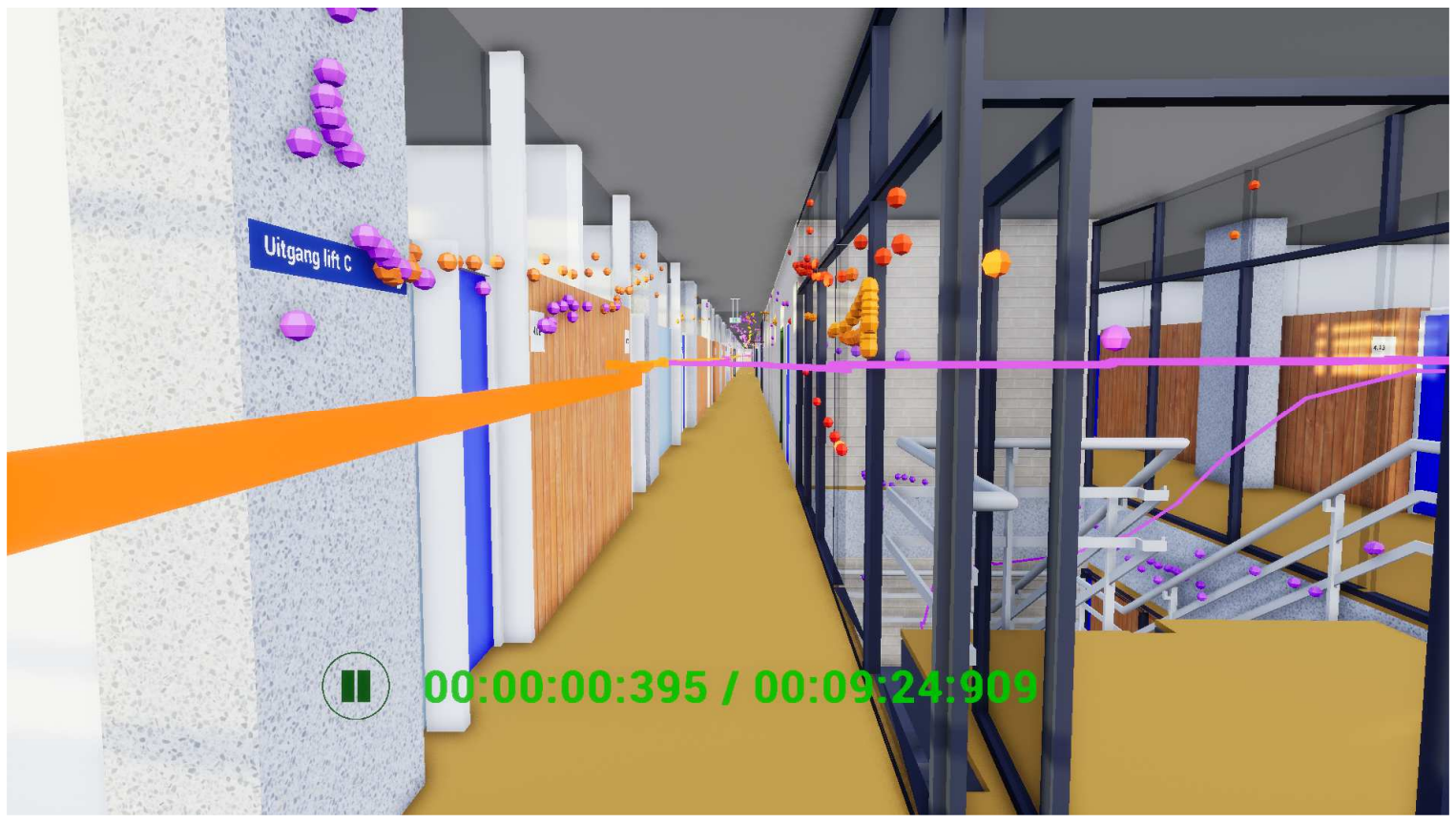

Figure 7: One example of the distribution of walking trajectories and gaze points in the virtual building

## 5 EVALUATION EXPERIMENT

In order to validate and ensure the correct functioning of the VR tool, a VR experiment has been designed and conducted. This section details the experimental design, the experimental apparatus adopted for this study, experimental procedure, data collection and participant's characteristics.

### 5.1 Experimental design

The experiment aims to test and evaluate the VR tool via investigating pedestrian route choice behaviour across 1) a horizontal level, 2) a vertical level, 3) a combination of horizontal and vertical level and 4) route and exit choice during evacuation. In total, the experiment entails two parts. The first part features walk-through assignments, and the second part features an evacuation experiment. Both assignments have no formal time limit. In accordance with the experiment description, participants consider all the information provided to them by the virtual environment and walk through the building.

The first part of the experiment is split into three distinct assignments featuring route choices across a combination of horizontal and vertical levels. Firstly, pedestrian route choice behaviour at the horizontal level is investigated. Participants are asked to find their way from Room 4.02 to Room 4.99 (Figure 1), which ensures they need to cross from the one main corridor to the other and walk the length of the building. Secondly, pedestrian route choice behaviour (including staircase choice) at the vertical level is investigated. Participants are asked to find their way from Room 4.99 to Room 2.01. This assignment requires participants to move between floors and walk the length of the building. Thirdly, pedestrian route and exit choice in both horizontal and vertical level are investigated. Participants are asked to find their way from Room 2.01 to Room 4.64, which forces them to switch floors, main corridors and walk through the building.

The second part of the experiment is to investigate pedestrian route and exit choice during evacuation. Participants are asked to evacuate from 4.64 and find an exit on the first floor (the exit floor underneath the second floor). When participants exit on the first floor, the experiment ends.

### 5.2 Experiment apparatus

Especially in a complex or large-scale virtual environment, immersion is one of the major key factors for being able to intuitively perceive all aspects of the scene [56]. In this experiment, participants were immersed in the virtual environment via a pair of earphone and the HTC Vive system, which consisted of a head-mounted display, one controller and two laser-based base stations. The UE4 and the SteamVR were used to run the virtual environment. All experiments were taken in a 3.4m x 2.5m room with a 2.5m high ceiling, lighted by fluorescent lighting, with no reflective surfaces and no exposure to natural lighting (Figure 8).

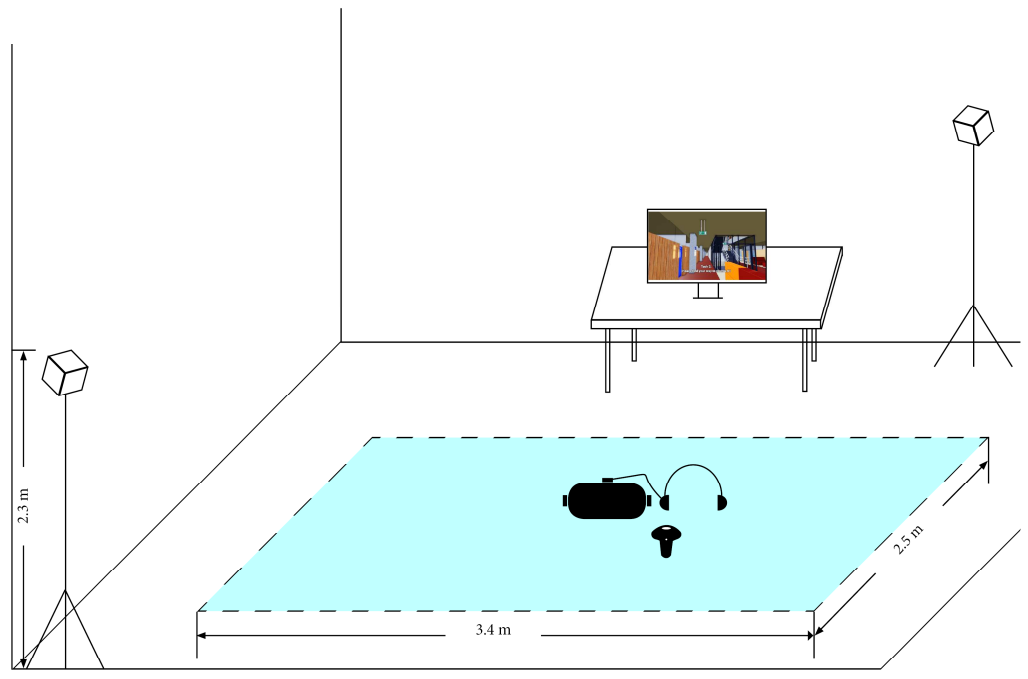

Figure 8: A simple illustration of the room setup

The HMD display has 360-degree head-tracking with a 110-degree field of view. It has two 3.4-inch RGB LCD screens, and each provides a resolution of 1080 x 1200 pixels (2160×1200 combined resolution) for 3D effects. It has a refresh rate of 90 Hz. Head tracking mechanisms translate movements of the participant's head into virtual camera movements [56]. Participants used one hand controller to move in the environment. Figure 9 shows one participant using the HMD display and one controller during the experiment. By simply holding the home pad of the controller, participants can move forward; by releasing the home pad, participants can stop moving. The direction of the movement is controlled by the orientation of the participant's head.

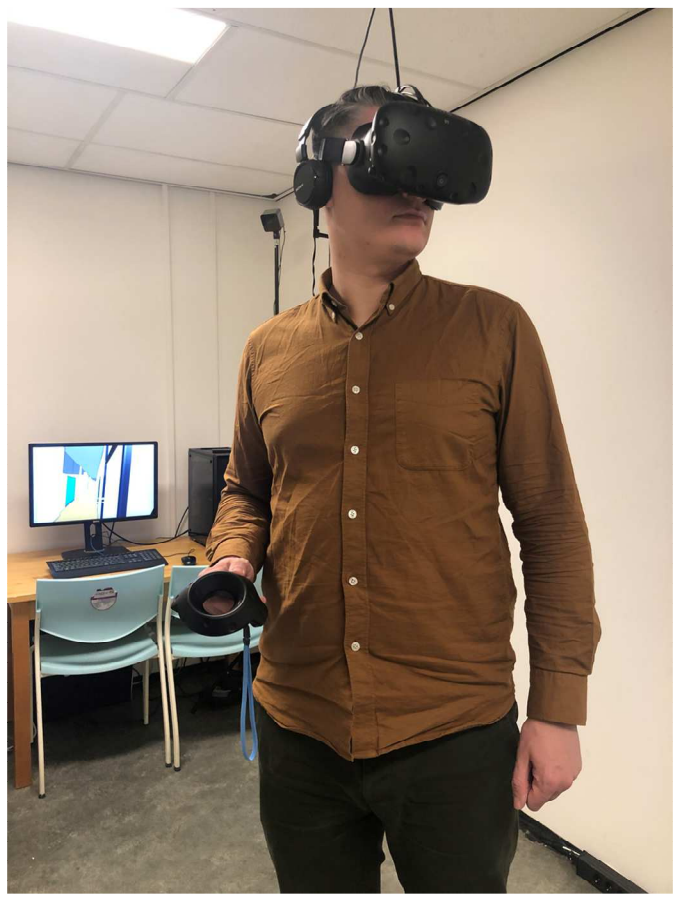

Figure 9: One participant was using the HMD display and hand controller during the VR experiment

HTC Vive provides a room-scale technology that allows the user to freely walk in real-life space and reflects their movement in the virtual environment. It is achieved by using tracking equipment, namely the base station (also called lighthouse). The base stations accurately track the position and orientation of the headset and the controller and translate this into the virtual environment in real-time. The base stations were replaced opposed to each other in the room with a 3.4m x 2.5m tracking area, which enables participants to move anywhere and re-orient themselves in any position within the range of the base stations. They were mounted on stable tripods at the height of 2.3m from the ground and were connected to each other via the sync cable. Once participants can move freely in the pre-defined area, it is necessary to protect them from running into the walls in the room. The measure here is showing participants the edge of the area when participants attempt to go beyond the tracking region.

In addition to the HTC Vive system, a pair of headphone was used by the participants. The headphone provided audio information to the participants and in case isolated the noise from the real-life environment.

### 5.3 Experiment procedure

The procedure of the VR experiment included the following parts, participants: 1) were introduced about the usage of the HMD and procedure of the experiment; 2) were familiarised with the test virtual environment and the HMD device in a simple training scenario; 3) took part in the official experiment; 4) filled in the questionnaire. Underneath, the four parts of the procedure are further explained. The VR experiment was approved by the Human Research Ethics Committee of the Delft University of Technology (Reference ID 944).

*1. Introduction*. Before the experiment, we made sure participants have normal sight or use corrective lenses. Once the participant arrived at the experiment room, the procedure of the experiment was introduced to the participant via a written instruction manual in order to ensure all participants had exactly the same information when entering the virtual environment.

*2. Familiarisation*. Participants were invited to wear the headset and headphone to walk through a test environment, which features a square area with obstacles randomly located in the area. Signs were added on the wall in the test environment. Participants were instructed to walk from A to B to C (Figure 10). This training assignment was used to familiarise the participants with the control of the device and discover any tendency of motion sickness. During the assignment, participants needed to perform basic movement operations and get acquainted with the system's mode of operation. The familiarisation phase lasted approximately 3 minutes. Participants who felt sick during this period were allowed to have a break, and after the break, they could decide whether to quit or continue the VR experiment.

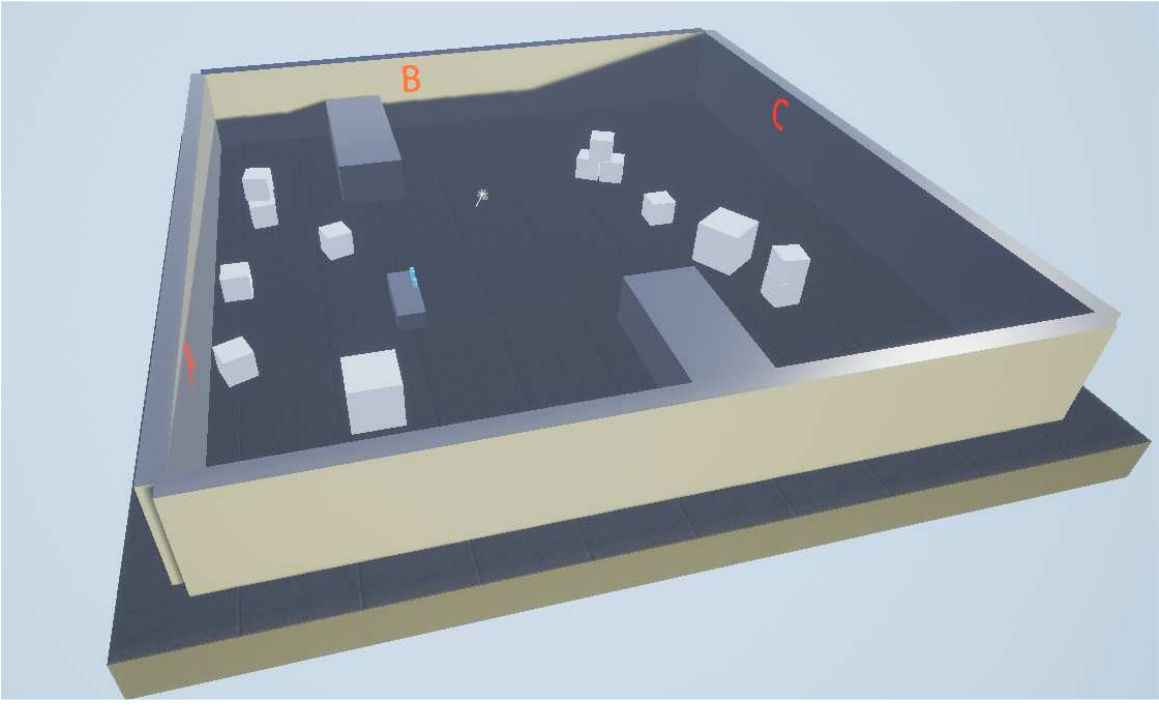

Figure 10: A screenshot of the test environment

*3. Performing the assignments*. After the familiarisation phase, participants were teleported to the actual virtual building. As stated in section 4.1, the start position is Room 4.02 (Figure 1), where participants were instructed to begin the first assignment. When participants reached the destination of the assignment, the information showed up to instruct participants to begin the next assignment (Figure 11). At the beginning of the fourth assignment, the evacuation alarm sound was automatically triggered, followed by a voice message instructing all people to evacuate from the building.

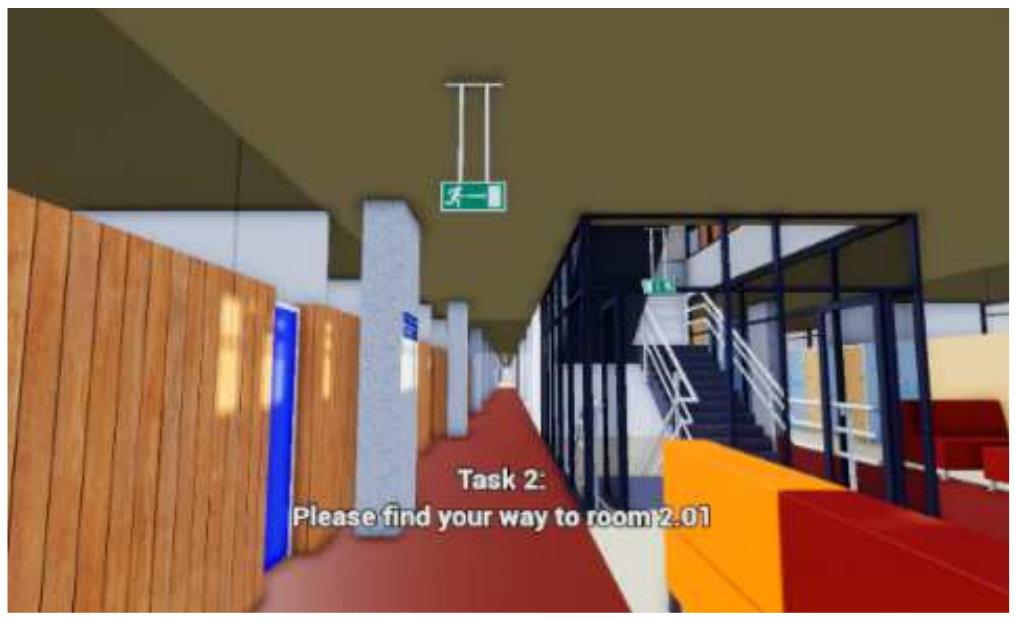

Figure 11: A screenshot of participant's view during the experiment, showing the current assignment

*4. Filling the questionnaire*. A questionnaire was provided to the participants directly after participants finished their assignments, which they filled in using a desktop located in the room. The last step of the procedure comprised of the researcher ensuring that participants were all right before they were allowed to leave the experimental room.

### 5.4 Data collection

The experiment collected two types of data, namely the behavioural data and questionnaire data. Firstly, participant behaviour in the virtual environment was recorded. Participant's positions, head rotations, gaze points, timestamp were recorded in milliseconds within the UE4. Jointly these collected data captured a rich set of information related to pedestrian route and exit choice behaviour. These data can be translated into four types of behavioural data, namely (1) participant route choices, (2) the travel time of participants during all four assignments, (3) the choice for the specific staircase(s) during each assignment and final exit choice, and (4) gaze points of how environmental elements draw the attention of participants.

Secondly, a questionnaire was designed in order to obtain the personal features and experiences of each participant regarding the virtual experiment. The questionnaire contained five sections: (1) participant's information, which included their socio-demographic information and their experience with VR and the experimental building in real-life, (2) the face validity questionnaire, which assessed the realism of the virtual environment, (3) the Simulator Sickness Questionnaire [57], which determined if participant's experience sickness throughout the experiment, (4) the System Usability Scale [58], which assessed the usability of the applied VR system as a pedestrian simulator, (5) the Presence Questionnaire [59], which measured participant's experience of presence in the virtual environment. Here, a comprehensive questionnaire was used to ensure that the authors are able to study the validity of the virtual environment and participant's VR experience in great detail.

### 5.5 Participant's characteristics

A total number of 37 participants took part in the VR experiment, and 36 participants finished all assignments. One participant asked to take a break during the third assignment and did not finish the whole experiment. Thus, the results discussed underneath are based on 36 participants, which included eighteen females and eighteen males. The age of these participants ranged from 17 to 41 years (M = 28.58 years, SD = 5.95 years). Table 1 presents the descriptive statistics of the participants. It shows that the participants were generally familiar with computer gaming and not very familiar with VR. Moreover, most of the participants received relatively high education. All participants were volunteers and experienced the experiment one by one. None of the participants complained about sickness during the experiment and all finished the experiment completely.

Table 1: Demographic information of participants.

| Descriptive information | Category | Number (percentage) |
|---|---|---|
| Highest education level | High school or equivalent | 5 (13.89%) |
| | Bachelor degree or equivalent | 6 (16.67%) |
| | Master degree or equivalent | 19 (52.78%) |
| | Doctoral degree or equivalent | 6 (16.67%) |
| Previous experience with VR | Never | 11 (30.56%) |
| | Seldom | 18 (50.00%) |
| | Sometimes | 6 (16.67%) |
| | Often | 1 (2.78%) |
| | Quite often | 0 (0.00%) |
| Familiarity with any computer gaming | Not at all familiar | 7 (19.44%) |
| | A-little familiar | 5 (13.89%) |
| | Moderately familiar | 8 (22.22%) |
| | Quite-a-bit familiar | 7 (19.44%) |
| | Very familiar | 9 (25.00%) |

## 6 RESULTS AND DISCUSSION

The main objective of the VR experiment is to demonstrate the functionality and evaluate the usability of the developed VR research tool. This section first illustrates the ability of the VR tool to collect pedestrian behavioural data, namely pedestrian route choice and exit choice behaviour, time spent, and point of interests. Next, this section examines the validity, realism and usability of the VR tool based on the results of the subjective data collected from the questionnaire (i.e., face validity, the simulation sickness, feeling of presence, system usability).

## 6.1 Objective standpoints

This sub-section presents an analysis of objective behavioural data collected during the VR experiment, including the time spent, the route and exit choice and point of interests.

### *6.1.1 Time spent on assignments*

One important measurement of pedestrian route and exit choice behaviour is the time spent on the assignments, which provides information regarding the ease with which participants navigate through the building. The time spent is defined as the time period between the moment in time a participant starts an assignment and the moment in time the participant arrives at the destination.

On average, participants spent 568.5 seconds (SD = 62.1 s) to finish four assignments. Figure 12 shows the distribution of the time spent by the participants on each assignment. On average, the participants spent the most time on assignment two (M = 200.5 s, SD = 18.5 s), followed by assignment one (M = 160.3 s, SD = 20.1 s), assignment three (M = 140.1 s, SD = 23.8 s), and the least time on assignment four (M = 66.2 s, SD = 11.4 s). This is in line with our expectation, as the minimum distance required to travel for each assignment also decreases in the same order. Besides that, we see that the time spent is clustered, in particular for assignments 2 and 4. This finding suggests that the variation in the walking speed was limited.

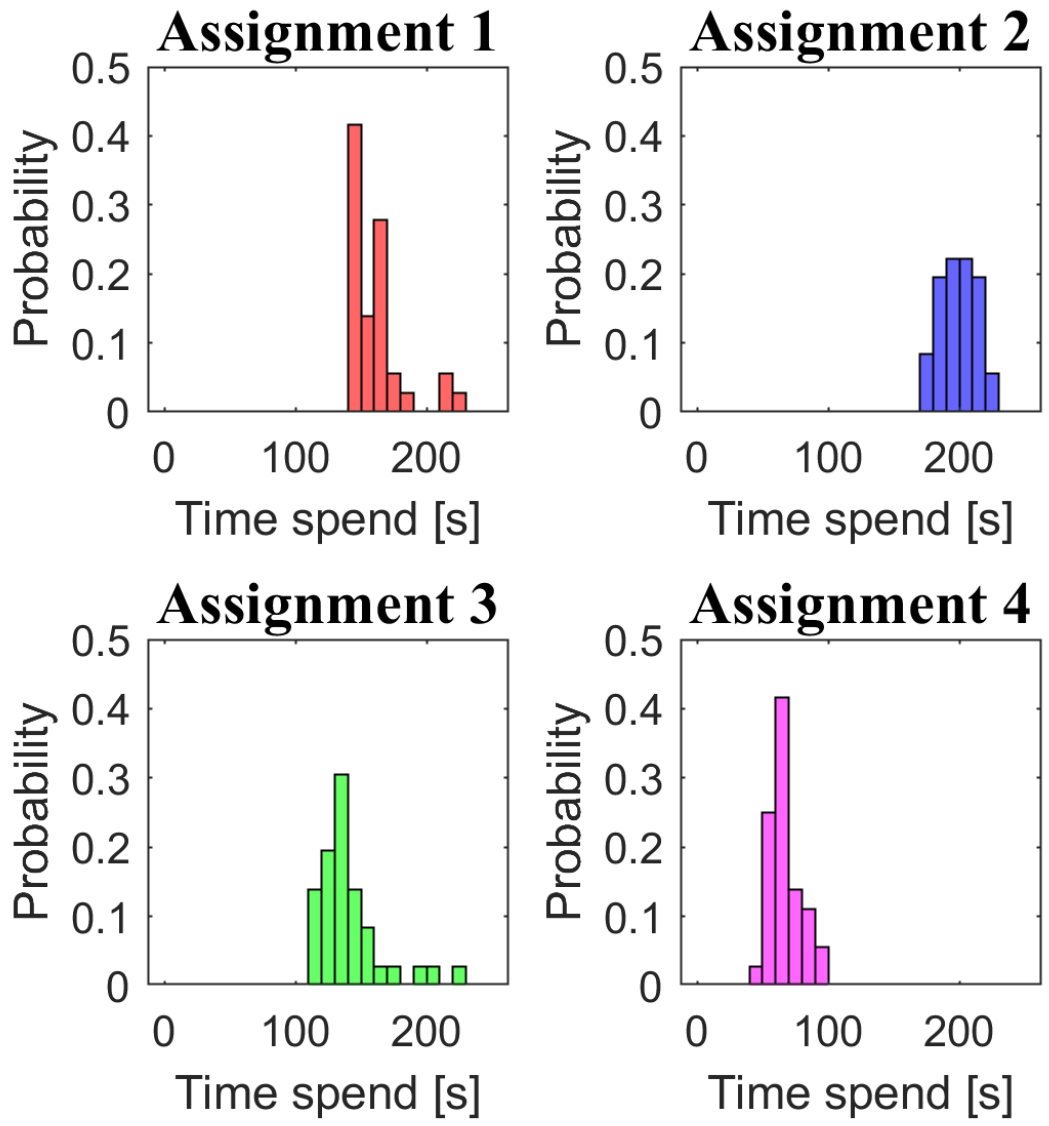

Figure 12: Histograms of the time spent by the participants

### *6.1.2 Route and exit choice behaviour*

The walking trajectories of participants are recorded at 10 Hz, which means that we obtain approximately 5600 data points per participant describing their movements throughout the four assignments. The trajectories clearly reveal the route choices of participants. For instance, Figure 13 shows the trajectory of one participant during all assignments. Here, the colour of the trajectory indicates the time: blue (start)-green-yellow-red (end). During the third assignment (Room 2.01 to Room 4.64), there was an obvious detour on the fourth floor when this participant clearly ‘missed’ the chances to use wider intersections to reach Room 4.64.

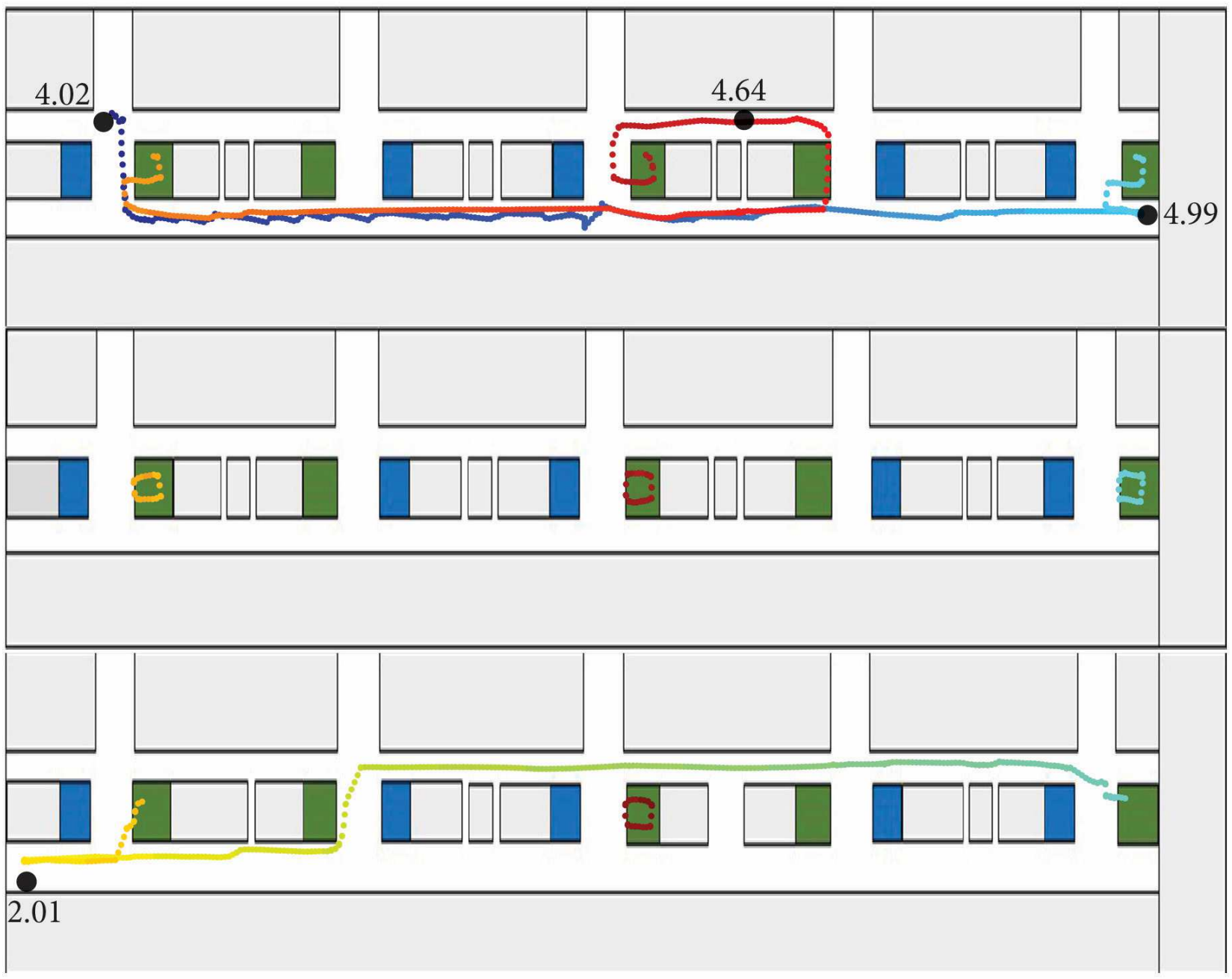

Figure 13: Visualisations of one participant's route choice during four assignments

Hölscher et al. (2007) [60] proposed three distinct strategies for pedestrian route choice in a multi-story building environment. Pedestrians use the central point strategy to find their way by frequently visiting well-known parts of the building, even if this requires considerable detours. Pedestrians that employ the direction strategy first move to the horizontal position of the destination as directly as possible (irrespective of level-changes). Pedestrians use the floor strategy first find their way to the floor of the destination (irrespective of the horizontal position of the destination). To analyse participants' route and choice behaviour, the complete set of walking trajectories was split into four separate sequences pertaining to each assignment. Figure 14 shows the walking trajectories of all participants during the first assignment (Room 4.02 to Room 4.99). Two major strategies regarding route choice were detected. Participants who adopted central point strategy first went straight from Room 4.02, then used the wider intersections to cross towards the other corridor, on which side Room 4.99 resides, and then continued walking towards the destination. There were also some participants that chose shortcuts and crossed the small intersections between the two major corridors. The other main strategy was direction strategy, participants chose to turn directly to other wider corridor at the first interaction and then walked straight down the corridor towards the destination. It is interesting to see, that participants clearly adopt different strategies in the virtual environment.

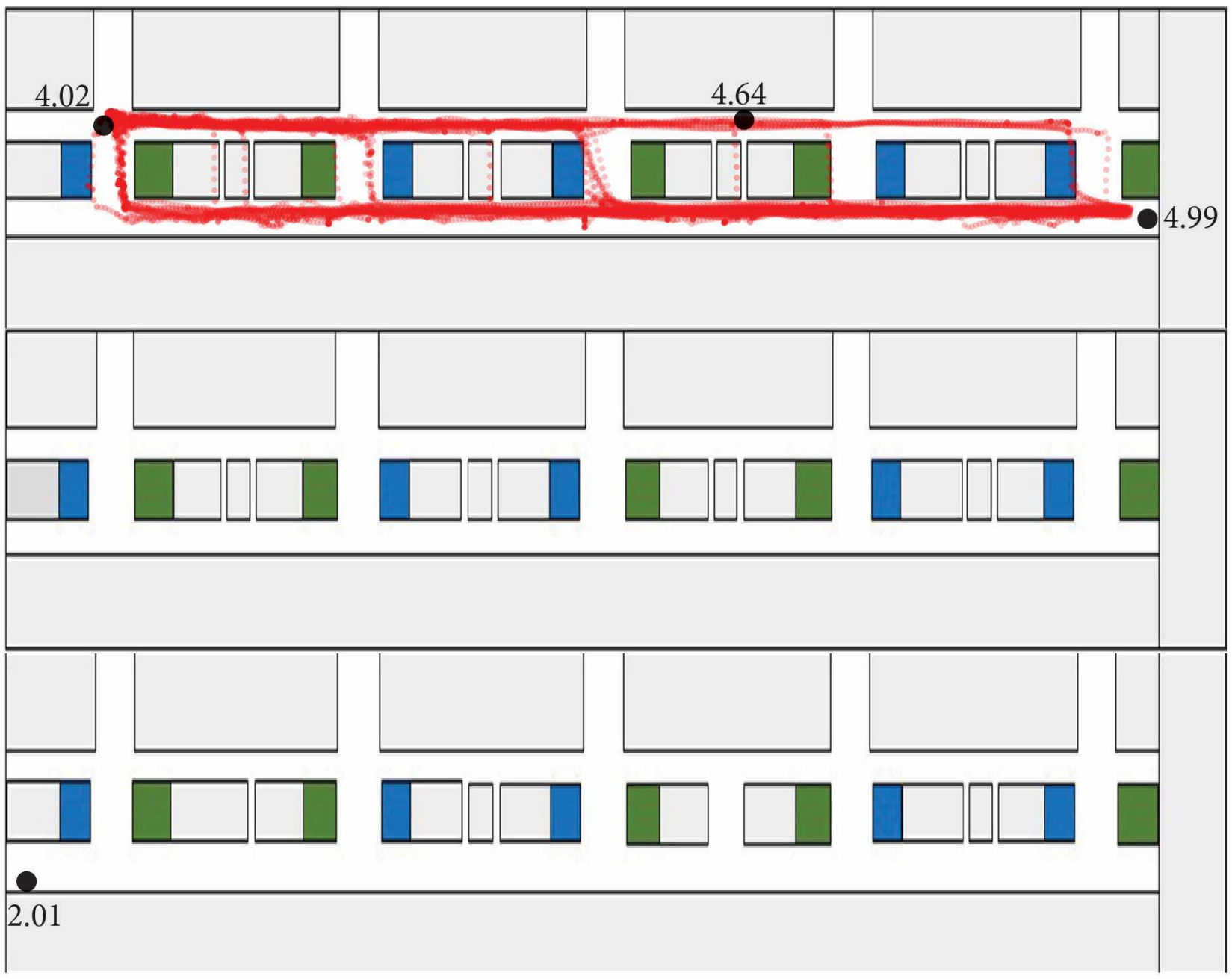

Figure 14: Participants' walking trajectories during the first assignment

After finishing assignment one, participants were instructed to find their way from Room 4.99 to Room 2.01. This assignment involved their movement across different floors in the building while both rooms remaining on the same side. Figure 15 shows the walking trajectories of all participants during this assignment. It shows that participants predominantly applied the direction and floor strategy. Participants who employed the direction strategy stayed on the fourth floor and walked to the direction of Room 2.01, then went down using one of the staircases. Participants who employed floor strategy initially went down to the second floor using the first staircase they encountered nearby Room 4.99, then they subsequently found the target room number within the floor. Only one participant chose to go from the fourth floor to the third floor, then to the second floor, a clear deviation from the other strategies.

After the first assignment, participants were clearly more aware of the structure of the building, for instance, rooms were located with even and uneven number on different sides. Thus, there was less switching of sides in their route because they were more conscious of Room 2.01 is located on the even side of the corridor. Only a few participants chose to stay on the opposite side of the building until they reached the last intersection point before their destination. It might because of the disorientation effects caused by changing floors [61]. Most participants chose to stay at the side of the building where Room 4.99 and 2.01 were located.

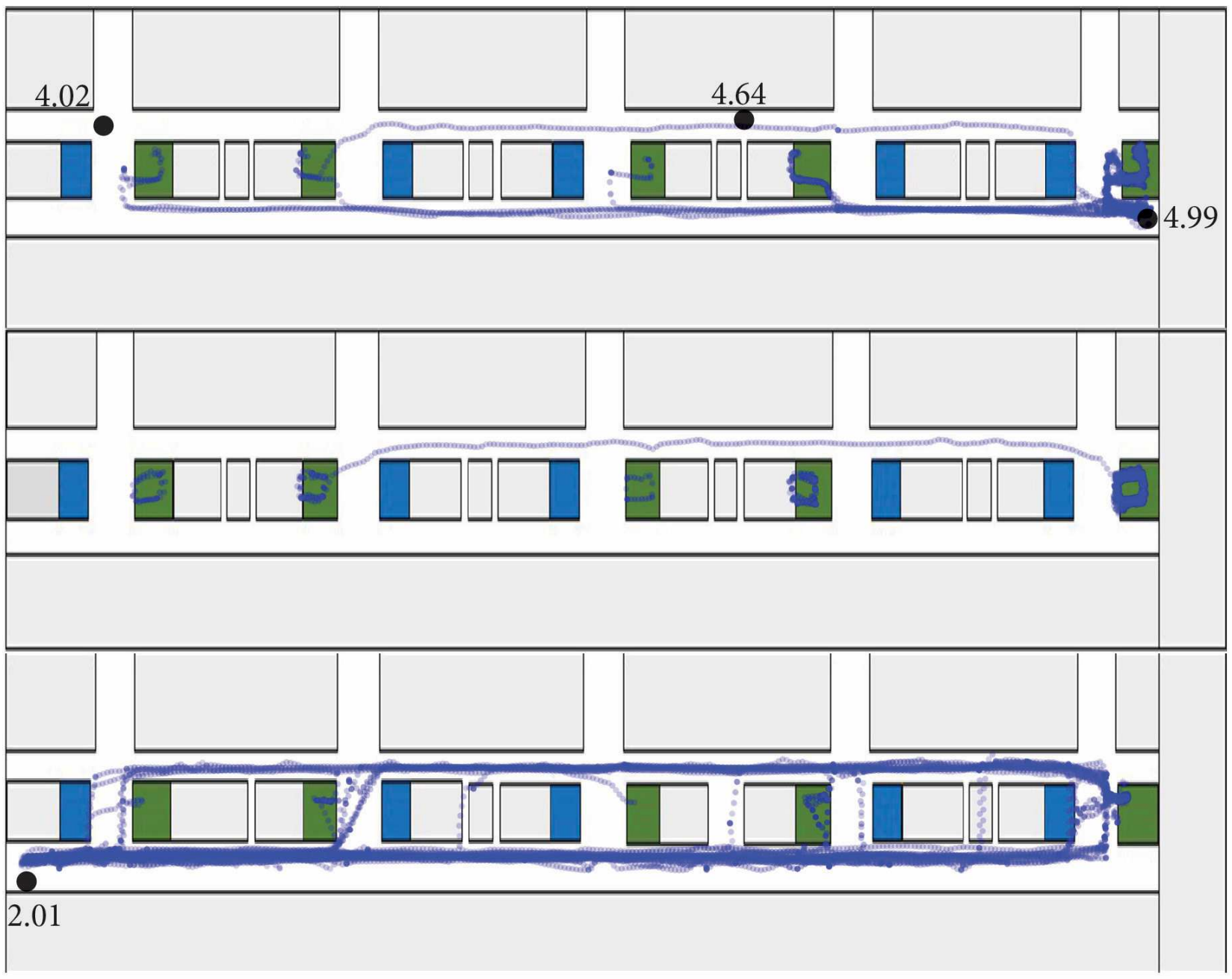

Figure 15: Walking trajectories during the second assignment.

Figure 16 shows the route trajectories of participants during the third assignment (Room 2.01 to Room 4.64). Participants predominantly employed the floor strategy who firstly climbed up the fourth floor using the first staircase they met and accordingly kept walking along the side of Room 4.64. Some others used direction strategy and first moved along the second floor before entering the stairs. Although it was the third assignment and some learning effect was recorded, for some participants it was still not entirely clear on which side of the destination would be. Hence, they adopted the central point strategy, stuck to the main corridor they already knew and took a detour via the other side of the building which is indicated by the green plots opposite Room 4.64.

Figure 17, furthermore, shows the aggregation of the trajectories of all participants during the last assignment (evacuate from Room 4.64 to an exit). Most of the participants either chose to go down from the right side of Room 4.64 or the left side. Even though five exits were available, only the exits near elevator C and D were chosen, which shows the usage of the exits is asymmetrical. This behaviour is in line with other studies look at exit usage [62–64]. Amongst the choices, 18 participants chose Exit C and 18 participants chose Exit D. These are the closet two exits for participants. This result is consistent with the studies which found that pedestrians were overall more likely to choose the nearest exits and shortest routes [16,29,30,65,66]. Interestingly, the split between the two exits is 50% - 50%, which suggests that we see no bias towards the side the participant approached Room 4.64 from. The results also suggest that participants prefer to use wider corridors among four assignments, which are concordant with studies of [35].

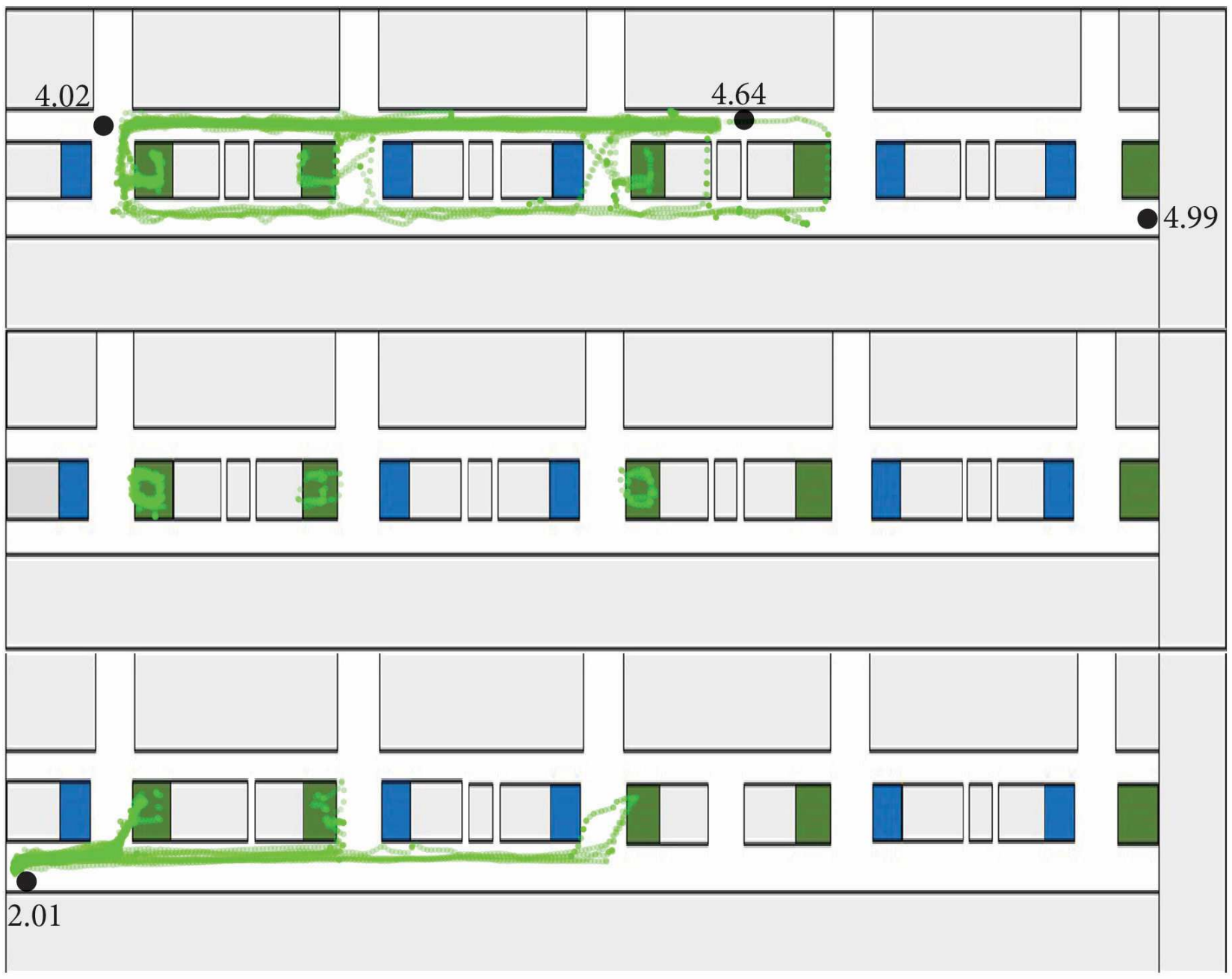

Figure 16: Walking trajectories during the third assignment.

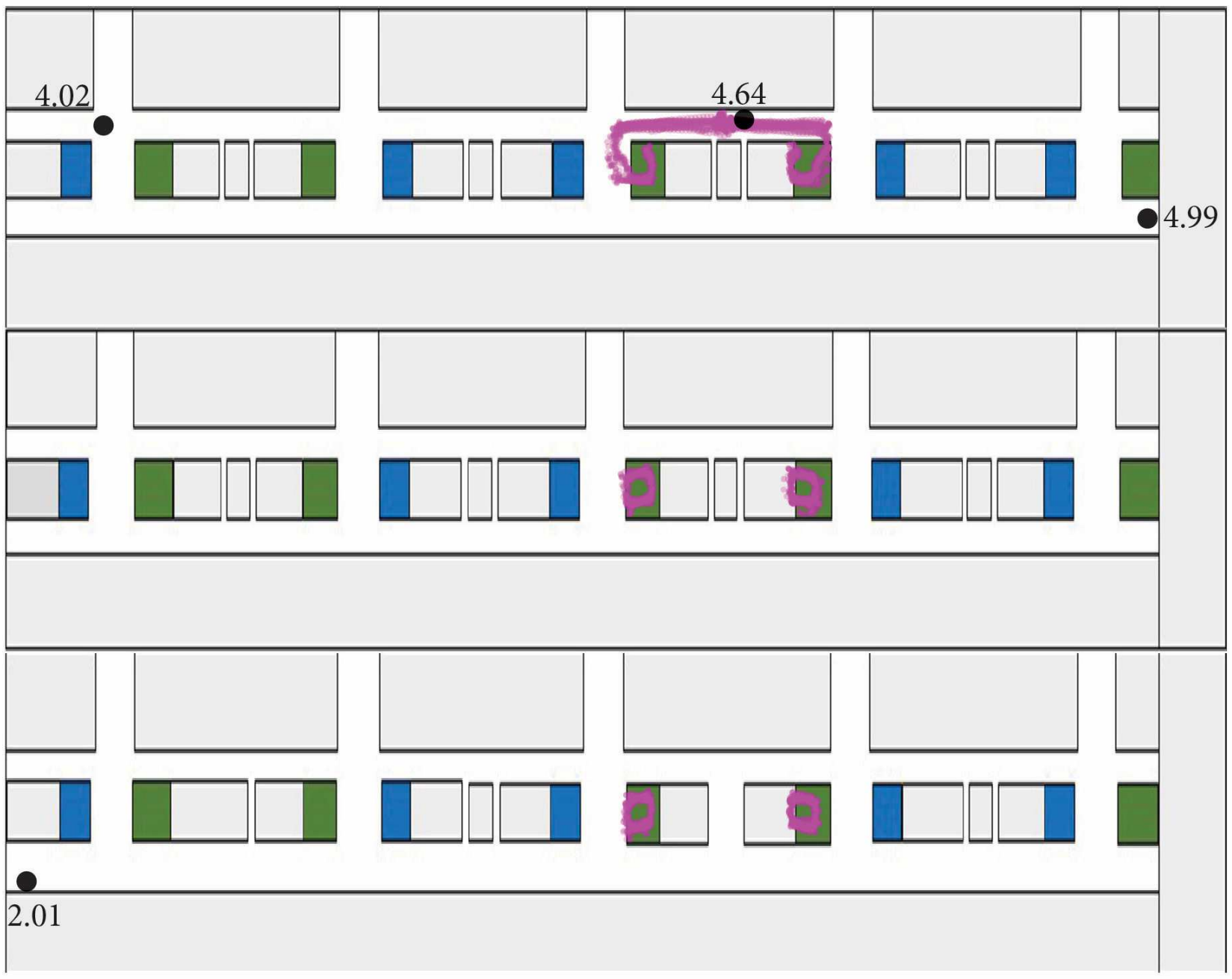

Figure 17: Walking trajectories during the fourth assignment

### *6.1.3 Point of interests in the space*

While performing the assignment, participants kept searching for information using the signs and landmarks on the route towards the destination. The collected gaze points data helps to analyse what objects or information in the virtual environment capture participants' attention. By combing the head rotations with route trajectory data, we can investigate how particular elements in space influence pedestrian route and exit choice behaviour. For instance, Figure 18 shows the scatter of one individual’s (the same participant in Figure 13) gaze points during all four assignments. The directionality of its gaze towards the side of the corridors indicates that the individual's focus has mainly been directed towards room number signs. Figure 19 shows the overall gaze points of all participants during four assignments. From this analysis, we can derive that during the first three assignments (Figure 19 a-c), the main visual attractions in the building are room numbers and fire doors (indicated by the red dots along with the room and red vertical lines across the main corridors). During the last evacuation assignment, participants paid more attention to the exit signs (Figure 19d), which are indicated by the red dots near the two staircases around Room 4.64.

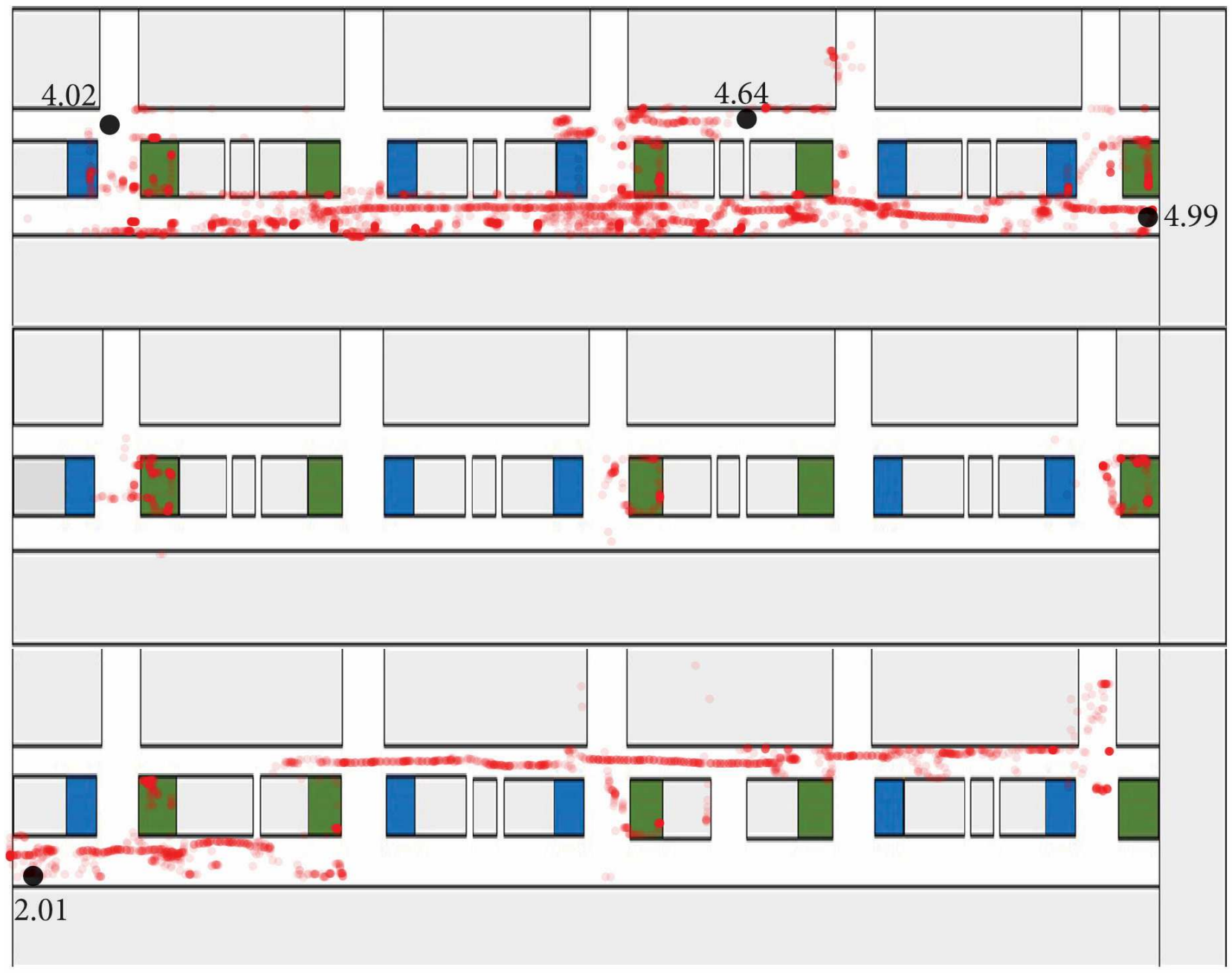

Figure 18: Visualisations of one participant's gaze points

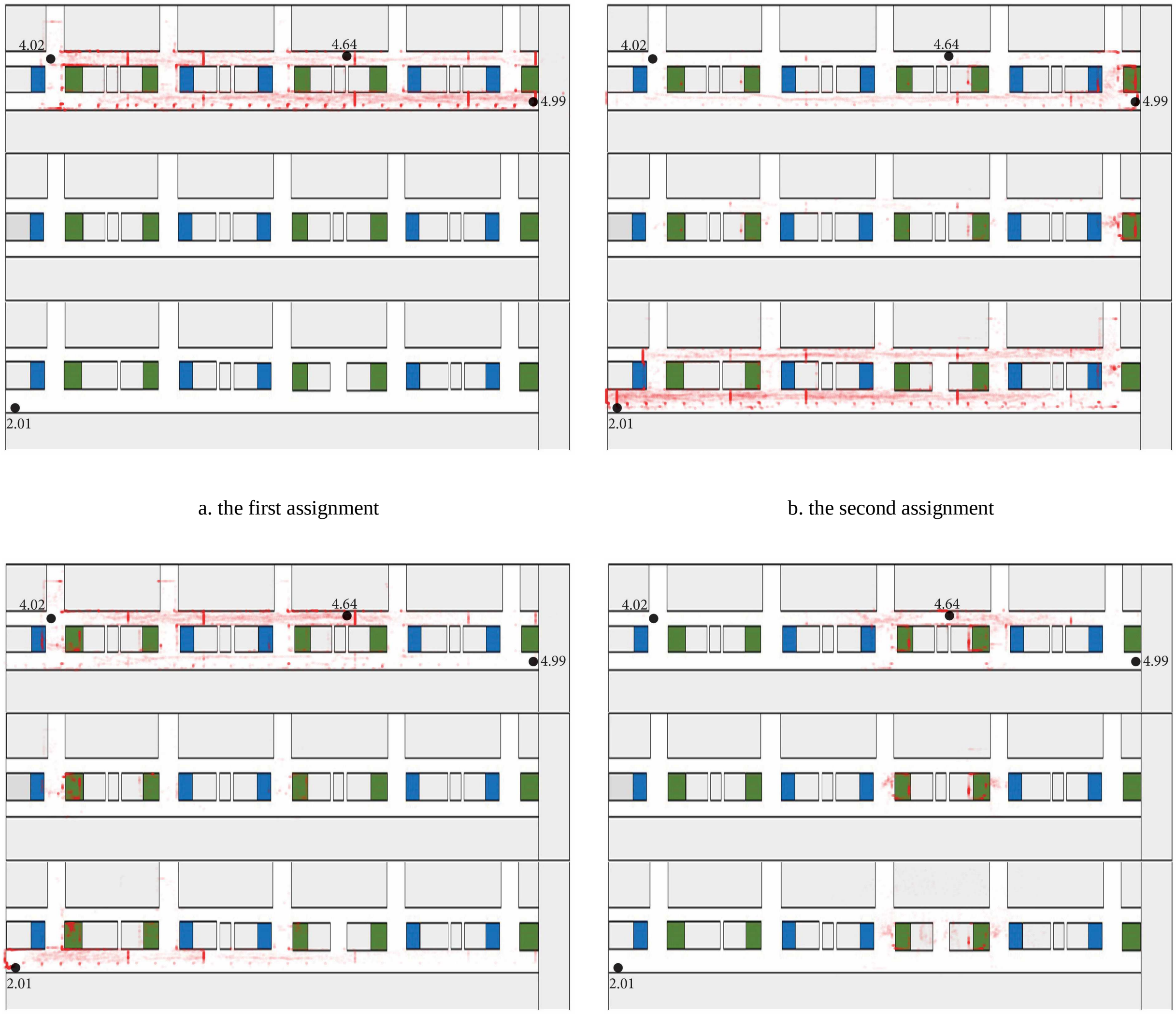

a. the first assignment

b. the second assignment

c. the third assignment

d. the fourth assignment

Figure 19: Visualisations of all participants' gaze points during four assignments

## 6.2 Subjective standpoints

This sub-section describes the results of subjective data derived by means of the questionnaires, namely the face validity questionnaire, the Simulation Sickness questionnaire, the Presence questionnaire and the System Usability Scale questionnaire.

### *6.2.1 Face validity*

Face validity refers to the degree of a simulator's realism compare to the real situation. The realism of participants' experience with the VR tool was assessed using a five-point Likert scale ranging from 1 (not at all realistic) to 5 (completely realistic). The results of the face validity are provided in Table 2. Amongst four elements regarding the realism of the virtual environment, the realism of the evacuation alarm sound received the highest score (M = 4.78), which shows participants were highly engaged in the assignment and felt the threaten of the emergency situation. The lowest score was assigned to the realism of the movement ability (M = 3.8). Participants needed to hold the controller's button while walking in the virtual environment; this might be the reason for having a lower score compared to other perspectives. Overall, the average score is 4.07, and three scores (out of four) are above 4, which indicate that the VR tool has a relatively high degree of realism with respect to the virtual environment and assignment design. Thus, these results establish the face validity of the VR tool.

Table 2: Rating of VR tool realism

| **The realism of the VR tool** | **Mean** | **SD** |
|---|---|---|
| The realism of the virtual building | 4.11 | 0.57 |
| The realism of the virtual furniture (chairs, doors, etc.) | 4.19 | 0.58 |
| The realism of the movement ability | 3.19 | 0.67 |
| The realism of the evacuation alarm sound | 4.78 | 0.42 |

### *6.2.2 Simulation sickness*

Simulation sickness is generally defined as the discomfort that arises from using simulated environments [67]. When designing a VR tool, it is essential to evaluate whether the tool potentially causes simulation sickness. The Simulator Sickness Questionnaire [57] determined participant's experience on a set of symptoms (e.g., fatigue, headache) in a 4-point Likert scale, from 0 (none) to 3 (severe). It can be summed to a total symptom score as well as grouped into three subscales, namely nausea, oculomotor disturbance, and disorientation. The total score is calculated by summing the reported values in each subscale and accordingly multiplying the result by 3.74 [68]. The total score of SSQ can range from 0 to 236. For each subscale, the scores are based on the reported scores for each symptom and then multiplied by the weight for that particular subscale.

The total score reflects the severity of the symptomatology of participants using the VR tool and index the troublesomeness of a simulator [69]. In the present study, the average total score of simulation sickness questionnaire is 13.61 (SD = 13.20) with up to a maximum of sixteen-minute exposure to the virtual environment. The total score is relatively low and only negligible symptoms or minimal symptoms were found among all participants according to the categorisation of symptoms [69]. Table 3 presents the results of each subscale of SSQ. It shows that the subscale of disorientation received the highest score, followed by Oculomotor and Nausea. Although disorientation subscale is related to vestibular disturbances such as dizziness and vertigo, high disorientation may be an indicator of having experienced higher levels of virtual presence [70]. The relatively high disorientation score might be the result of rotation-induced effects. That is, while participants walking through the virtual environment, they can rotate their head side to side, which might cause a response lag. Moreover, the current experiment assignments involved changing floors and some turning movements on the stairs in the virtual building, which are key sources of disorientation about one’s heading and position in a building. The relation between disorientation and floor changes was also found in [61].

Table 3: Subscales of SSQ: Means and standard deviations.

| **Subscale** | **Mean** | **SD** |
|---|---|---|
| Nausea | 7.95 | 11.06 |
| Oculomotor | 13.27 | 12.07 |
| Disorientation | 14.69 | 18.81 |

### *6.2.3 Feeling of presence*

The sense of presence reported by participants is a key factor to evaluate the effectiveness of virtual environments [59]. The Presence Questionnaire (PQ) is a wildly applied questionnaire to measure the degree of participant's feeling of presence in the virtual environment. It depends on four subscales, namely Sensory fidelity, Immersion, Involvement and Interface quality [59]. Participants use a 7-point scale format to rate 29 questions.

The total score was counted by summing the reported scores of the 29 items. The average total score for PQ in this study is 146.17 (SD = 13.56), which indicates that the participants had a strong sense of presence. Table 4 shows the mean value and standard deviations of the four subscales in the PQ questionnaire. The Immersion subscale received the highest score, which confirms that the participants felt a high level of immersion in the designed virtual environment. Meanwhile, the score of Sensory fidelity established the accuracy of sensory stimulation. The Involvement score indicates that participants were able to focus their attention and energy in the virtual environment. The interface quality score shows that control devices had little distraction from participants' assignment performance, and the participants were able to concentrate on the assignments. Furthermore, participants' response to Question 8, (i.e., “How much did your experience in the virtual environment seem consistent with your real-world experience ?”, M = 5.13, SD = 0.96) indicates that the experiences in the virtual building are consistent with the real-world experience.

Table 4: Subscales of PQ: Means and standard deviations (range from 1 to 7).

| | **Involvement** | **Sensory fidelity** | **Immersion** | **Interface quality**[a] |
|---|---|---|---|---|
| Mean | 4.84 | 4.94 | 5.76 | 4.14 |
| SD | 0.61 | 0.85 | 0.49 | 0.97 |

[a] Reversed items

### *6.2.4 Usability of the VR tool*

System Usability Scale questionnaire represents a composite measure of the overall usability of the simulator system [58]. It contains questions such as, "I thought the system was easy to use" and "I found the various functions in this system were well integrated". Participants rated the ten items of this questionnaire on a 5-point Likert scale (i.e., 1 = strongly disagree, 5 = strongly agree). The total score of SUS is calculated by summing the converted responses on ten items and accordingly multiplying the result by 2.5. The total score of SUS ranges from 0 to 100.

The total score of SUS can be translated into adjective ratings for interpreting the results, such as 'worst imaginable', 'poor', 'OK', 'good', 'excellent', 'best imaginable' [71]. In the present study, the average score of the VR tool is 81.32 (SD = 12.12), which suggested 'excellent' usability of the VR simulator. Meanwhile, according to the observation from the researcher, all participants were able to easily understand all four assignments and how to use the HMD and controller to interact with the virtual environment.

## 7 CONCLUSIONS AND FUTURE RESEARCH

This study presents a new VR tool that is developed by means of Maya and UE4 which provides substantial benefits for collect pedestrian route and exit choice behavioural data in a multi-story complex building. The VR tool supports free movements in all directions, allows for the monitoring of pedestrian behaviour throughout a complex building and includes the tracking of walking trajectories, gaze points and head movements. The VR tool addresses several limitations with respect to using VR for pedestrian behaviour research, such as free movement across horizontal and vertical level, the accurate collection of a wide range of data related to pedestrian behaviour.

This primary experiment aimed to demonstrate the functionality of the VR tool by showcasing participant's time spent, route and exit choice and point of interests in the space. Additionally, the realism of participants' experience with the VR tool, simulation sickness, the feeling of presence in the virtual environment and the usability of the VR tool were evaluated via questionnaires.

The results of this study show that this VR tool is able to collect participants’ route and exit choice behavioural data. It shows that participants predominantly applied the direction and floor strategy when walking in a multi-story building environment. Under evacuation conditions, participants were overall more likely to choose the nearest exits and shortest routes. Analysis of the gaze points shows that room numbers, fire doors and exit signs were the major attractors in the building. The questionnaire data showed that the VR tool had a high degree of realism and participants experienced a good feeling of presence in the virtual environment. Participants also only experienced minimal symptoms during the experiment. Meanwhile, Participants gave the simulator good marks on usability. To summarise, the results of this study confirm that this new VR tool is capable of precise collection of data pertaining the movement and choice behaviour of pedestrians in a complex multi-story building and the show that the high realism of the virtual environment, high immersive feeling, high usability, and low simulator sickness incidence.

Despite the innovation presented by the VR tool, this study has some limitations and provides implications for future research. First, the current applied HMD device only uses head tracking to present participants movements in the environment. In future research, applying other sensors, such as eye-tracking and body-tracking, would allow researchers to track pedestrian gaze points and movements more precisely. Second, although the face validity shows the high level of realism of the developed virtual environment, further validation studies are needed to determine whether the pedestrian movement and choice behaviour, gaze points are also similar to pedestrians’ behaviour in real-life. Third, no other agents or socially relevant variables were added at this stage because the goal of the current VR tool was to investigate the interaction between pedestrian and environment. One of the advantages of VR is the ability to rapidly change the scenario or add other elements to the virtual environment. The researchers of this study are continuing to explore the use of VR in other perspectives of pedestrian route and exit choice behaviour, for example, adding multiple users in the environment simultaneously, wayfinding information or dynamic obstacles and investigate their influences on pedestrian behaviour.

## ACKNOWLEDGMENTS

We thank the China Scholarship Council for their financial contribution. This research is also supported by the ALLEGRO project, which is financed by the European Research Council (Grant Agreement No. 669792). We also thank Arno Freeke and Arend-Jan Krooneman from the VR Zone of the Delft University of Technology for the help with developing the VR tool.